# Rethinking the Foundations of the Theory of Special Relativity: Stellar Aberration and the Fizeau Experiment

A.F. Maers[1] and R. Wayne[2]

[1]Department of Biological Statistics and Computational Biology, Cornell University, Ithaca, NY 14853 USA
[2]Department of Plant Biology, Cornell University, Ithaca, NY14853 USA

In a previous paper published in this journal, we described a new relativistic wave equation that accounts for the propagation of light from a source to an observer in two different inertial frames. This equation, which is based on the primacy of the Doppler effect, can account for the relativity of simultaneity and the observation that charged particles cannot exceed the speed of light. In contrast to the Special Theory of Relativity, it does so without the necessity of introducing the relativity of space and time. Here we show that the new relativistic wave equation based on the primacy of the Doppler effect is quantitatively more accurate than the standard theory based on the Fresnel drag coefficient or the relativity of space and time in accounting for the results of Fizeau's experiment on the optics of moving media—the very experiment that Einstein considered to be *"a crucial test in favour of the theory of relativity."*

The new relativistic wave equation quantitatively describes other observations involving the optics of moving bodies, including stellar aberration and the null results of the Michelson-Morley experiment. In this paper, we propose an experiment to test the influence of the refractive index on the interference fringe shift generated by moving media. The Special Theory of Relativity, which is based on the relativity of space and time, and the new relativistic wave equation, which is based on the primacy of the Doppler effect, make different predictions concerning the influence of the refractive index on the optics of moving media.

1. Introduction

Albert Einstein related to R. S. Shankland [1] on February 4, 1950 that the observational results of stellar aberration and Fizeau's experimental results on the speed of light in moving water "*were enough*" for him to develop the Special Theory of Relativity, which states that the difference in the observations made by an observer at rest with respect to the source of light and the observations made by an observer moving with respect to the light source is a consequence only of the relativity of space and time. In fact, Einstein [2] wrote that the Fizeau experiment, which could be viewed as a determination of the correct relativistic formula for the addition of velocities and which showed that the simple Galilean addition law for velocities was incorrect, was "*a crucial test in favour of the theory of relativity.*"

In this introduction, we provide context for Hippolyte Fizeau's celebrated experiment on the optics of moving media by recounting the observations, experiments, mathematical derivations and interpretations concerning stellar aberration that led up to Fizeau's experiment, and its subsequent interpretation in terms of the Special Theory of Relativity. While this pedagogical tack involves a discussion of the complicated, contentious, and contradictory mechanical properties of the 19th century aether, we want to emphasize at the onset that we have no intention of slipping such an aether back into modern physics. In the Results and Discussion section, we present a meta-analysis that shows that the new relativistic wave equation based on the Doppler effect is quantitatively more accurate than the standard theory in accounting for the results of the original and replicated versions of the Fizeau experiment concerning the optics of moving media. We also show that stellar aberration is mathematically related to the new relativistic Doppler effect through the angular derivative.

The phenomenon of stellar aberration, which was so important for the development of the Special Theory of Relativity, was serendipitously discovered by James Bradley [3,4,5], who in his unsuccessful attempt to observe stellar parallax in his quest to provide evidence for the Copernican heliocentric universe, noticed that he had to tilt his telescope in the direction of the movement of the Earth (Figure 1) in order to see the bright star named γ Draconis in the constellation Draco, which is almost perpendicular to the elliptic path the Earth takes in its annual revolution around the sun. Bradley discovered that the position of the fixed star was not correlated with the change in the *position* of the Earth in its annual voyage around the sun, as would be expected from Robert Hooke's [5,6] previous observations of stellar parallax, but with its annual change in *velocity*.

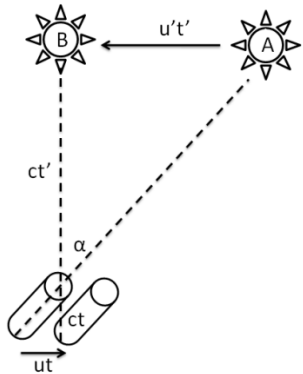

Figure 1. The aberration of starlight that results from the relative motion between the star and the observer on Earth. The star emitted light in the past that will form the image observed in the present. The time delay is due to the finite speed of light. Stellar aberration arises from the finiteness of the speed of light and there would be no stellar aberration if the speed of light were infinite and the light from the star formed an image without delay. In navigational terms, the "past" position of a star (A) is analogous to its apparent position at the present time and the "present" position of a star (B) is analogous to its true position. While the apparent position of the star is relatively easy to consider as an instantaneous image, determining the true position of the star at the present instant of time, requires taking a number of significant physical phenomena into consideration, including the relative velocity of the Earth, the position of the observer, the exact time of day and the refraction of the atmosphere. The magnitude of the aberration, which is given by the aberration angle (α), depends on the ratio of the relative velocity of the star and the telescope to the speed of light (c). For small angles, $\alpha = \tan \alpha = \frac{u'}{c} = \frac{u}{c}$ where $u'$ is the velocity of the star relative to a stationary observer and $u$ is the velocity of the Earth relative to the fixed star.

Bradley, who was a proponent of the corpuscular theory of light, explained this new motion of the fixed stars by assuming that the particles of light from a fixed star had to enter the front lens of a telescope and pass through the telescope to the eyepiece while the telescope was moving. If the telescope were at rest with respect to the star, then one would point the telescope directly at the star, almost perpendicular to the ecliptic. However, since the telescope was on the Earth, which was moving around the sun with a speed approximately equal to 30 km/s ($\approx$ 2πAU/year), then one had to tip the telescope downward in the direction of motion in order to see the star through the eyepiece. The tipping angle would allow the bottom of the telescope to lag behind the top of the telescope so that the light particles would travel down the telescope barrel without hitting the sides. This phenomenon is known as stellar aberration [7,8,9,10,11,12,13,14,15,16], and the angle that prescribes the difference between the observed position of the fixed star and the actual position at the instant of observation, is known as the angle of aberration ($\alpha$). The average angle of aberration is approximately 20 seconds of arc ($\approx 10^{-4}$ radians), and it is a result of the movement of the Earth and the finite speed of light. According to Bradley [3], the aberration *"proceeded from the progressive motion of light and the Earth's annual motion in its orbit. For I perceived, that, if light was propagated in time, the apparent place of a fixed object would not be the same when the eye is at rest, as when it is moving in any other direction, than that of the line passing through the eye and object; and that when the eye is moving different directions, the apparent place of the object would be different."* The angle of aberration relates the position of the star in the past at the instant when it emitted the light that will form the image, to the position of the star in the present at the instant of time when the image is observed. The tangent of the angle of aberration is equal to the ratio of the velocity of the Earth ($u$) to the velocity of light ($c$):

$$\tan \alpha = \frac{u}{c} \quad (1)$$

From the angle of aberration and the velocity of the Earth's motion, Bradley calculated that it would take eight minutes twelve seconds for light to propagate from the sun to the Earth. This means that at the present time, we see an image of the sun as it was in the past. We would like to emphasize the fact that, as a consequence of the finite speed of progression of light [17,18,19,20], a live image at the present represents the object in the past[1]—a truism first put

---

[1] The light that forms the image of *Eta Centauri* today was emitted over 300 years ago, just before Bradley discovered stellar aberration. Makena Mason wrote a poem for Bio G 450 (Light and Video Microscopy at Cornell University) that emphasizes the time it takes light to propagate:

The act of observing
Photons moving particles
The present never seen.

forward by Empedocles, discussed by Galileo, Cassini, Roemer, Fermat, Huygens, Newton, and deeply appreciated by Bradley.

Pierre-Simon Laplace[2], who was also a proponent of the corpuscular theory of light, hypothesized that, as a consequence of gravitational attraction between the mass of a star and the corpuscle of light, the more massive the star, the slower the light that emanated from it would be. Laplace requested that Dominique-François Arago undertake a study of the aberration of starlight in order to investigate the effect of the Earth's motion on the velocity of light emitted by the various stars. Arago reckoned that the refractive index of a glass prism depended in some way on the ratio of the speed of light in air to the speed of light through the glass, and he hypothesized that the daily and annular motion of the Earth would either add to or subtract from the components of the velocities of starlight parallel to the Earth's motion and thus change the refractive index of a glass prism. Accordingly, Arago reasoned that the angle of refraction given by the Snell-Descartes Law would vary with the motion of the Earth, and as a result, the angle of aberration measured through a glass prism should also vary with the motion of the Earth (Figure 2). However, around 1810, when Arago made the observations, he found, contrary to expectations, that within experimental error, a glass prism introduced to the front of his telescope refracted the starlight the same amount independent of the motion of the Earth and thus had no effect on the observed aberration of starlight [7,22,23,24,25,26].

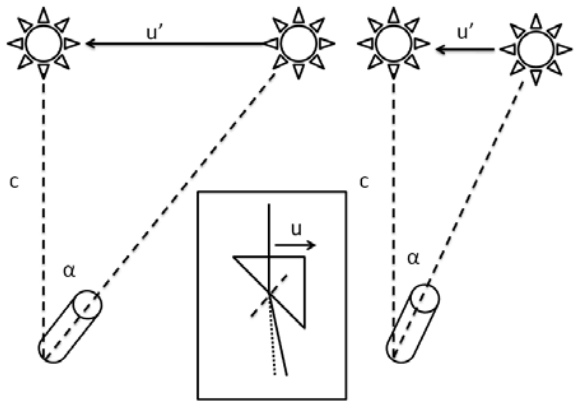

Figure 2. The increase or decrease in the angle of aberration ($\alpha$) expected by Arago as a result of the differing values of the relative motion ($u'$ or $u$) between the Earth and the star. The inset shows the predicted change in the angle of refraction caused by a glass prism as a result of the Earth's motion assuming that the index of refraction varies in a velocity-dependent manner $(\frac{3}{2} \frac{(1+\frac{u}{c})}{(1+\frac{u}{v})})$ and $\frac{3}{2}$ is the refractive index of glass at rest. The solid line gives the refracted ray when $u = 0$, and the dotted line gives the refracted ray when $u > 0$ and the refractive index decreases.

---

[2] Laplace believed that it was unreasonable to assume that any force, including the gravitational force, propagated from the source instantaneously [21].

In order to test the effect of the motion of the Earth on the refraction of light, Arago made his astronomical observations with or without a glass prism in front of the telescope both in the spring and in the autumn at 6 AM and 6 PM. He made the observations on aberration when the Earth was moving in opposite directions relative to the fixed stars so that the range of velocities of starlight would be the greatest. Nevertheless, Arago found that the angle that the starlight was refracted by the achromatic glass prism was constant, within experimental error, and independent of the velocity of the Earth.

Arago based the angle of incidence on the apparent position of the star that he observed at a given time, and he determined the angle of aberration from the difference between the apparent position of the star at the instant of observation and the actual position at the instant he observed the image. The apparent position was the position the star held at an instant of time in the past when it emitted the light that formed the image observed by Arago at a later instant of time. The finite speed of light meant, by necessity, that the image of the star was not formed instantaneously and simultaneously with the emission of the light that would later form the image of the star. Arago's null result meant that while the Snell-Descartes Law held for the refraction of light when the source, glass prism, and observer were all at rest relative to each other, it did not hold when one took into consideration the velocities of the source, glass prism, and observer from *any* inertial frame of reference as would be expected from Galilean relativity [27]. How could two optical phenomena have such conflicting dependencies on the velocity of the Earth? Stellar aberration was a result of the Earth's motion while refraction was independent of the Earth's motion. According to Arago, who at the time was a proponent of the corpuscular theory of light, the lack of effect of the glass prism on the aberration angle could be explained if each star emitted light with a wide range of velocities but the human eye could only observe light traveling within a narrow range of velocities. Consequently, it appeared that the limitations of the human eye were responsible for the null result. This was a reasonable interpretation given the then recent discoveries by William Herschel and Johann Ritter of invisible heat (infrared) rays and chemical (ultraviolet) rays on either end of the visible spectrum.

While Bradley and Arago considered light to consist of particles, Thomas Young [28] thought that the aberration of starlight could be reconciled with Robert Hooke's and Christiaan Huygens' recrudescent wave theory of light if "*the luminous aether* [which would solely set the speed of light] *pervades the substance of all material bodies with little or no resistance, as freely perhaps as the wind passes through a grove of trees."* While the wave theory could account for stellar aberration, it was unable to account for the null effect observed by Arago about six years after Young wrote these words. Arago, who was unhappy with his own explanation of the null result, asked Augustin-Jean Fresnel if he could come up with an additional hypothesis that could reconcile the null result with the wave theory of light. Since the mechanical wave theory of light, unlike the corpuscular theory of light, required a luminiferous aether, perhaps a reasonable hypothesis concerning a mechanical property of the aether would account for the null effect. Fresnel realized that if the Earth transmitted its total motion to the aether, then the Snell-Descartes Law of refraction would hold and Arago's results would be easy to understand because a glass prism would refract light the same way no matter what the velocity of the Earth was. However, an aether with this property would make the phenomenon of aberration of the fixed stars, impossible. By contrast, while a stationary aether would allow the phenomenon of aberration of the fixed stars, it would result in a velocity dependence of the Snell-Descartes Law.

Fresnel needed a way to reconcile these two mutually irreconcilable properties of the aether. He deduced that the aether could be endowed with a property that would permit the observed stellar aberration while still allowing it to share in part the velocity of the Earth. Such an aether would allow the starlight moving through a transparent medium to be pushed or pulled from its position predicted by the Snell-Descartes Law to its observed position, making such a null effect intelligible.

Fresnel [29,30,31,32] proposed a physically plausible mechanism based on the nascent mechanical wave theory that was able to quantitatively account for Arago's null result. According to Fresnel's mechanical wave theory, the square of the speed of light was inversely proportional to the density of the aether, and since according to the wave theory, the speed of light was slower in a glass prism than in the vacuum, the density of the aether in the glass prism would be greater than the density of the aether in the vacuum. Fresnel postulated that a moving glass prism did not carry all of its aether along with it, but only the part that is in excess relative to the vacuum. Consequently, the speed of light propagating through the moving glass prism, which was a function of the density of the aether, would be a weighted average of the speed of light through the stationary aether and the speed of light through a stationary glass prism. The weighting factor that characterized the proportion of aether carried along by the glass prism moving at velocity $u$, would be $\phi$. Consequently, the aether within the prism would move at weighted average velocity $u\phi$ where $\phi$ became a function to be determined that would, by necessity, quantitatively lead to the null result.

When modeling velocities, Fresnel had to take into consideration that all velocities are relative and must be designated with respect to a reference frame that can be operationally defined as static. Since the Earth rotates around its axis and revolves around the sun, it certainly is not a static reference frame; however, it does serve as a convenient, single reference frame for characterizing simply the motion of stars relative to an observer at rest with respect to the Earth. By applying the somewhat tedious but reliable techniques used in navigation for characterizing space and time, an observer at any location on the Earth can intelligibly describe the "present" position of the star to an observer anywhere else on Earth. On the other hand, a static aether, like the one put forward by Young, would provide an ideal single frame of reference for characterizing velocities (Figure 3). While a frame of reference can be arbitrarily chosen in order to provide the simplest mathematical formulation of the physical events in question, a true law of nature should not be restricted to the esoteric properties of one frame of reference but should include the necessary transformations so that the law is applicable to an observer in any inertial reference frame who is making measurements of the events in question. Fresnel first considered the observation of starlight by an observer, such as Arago, at rest with respect to the Earth.

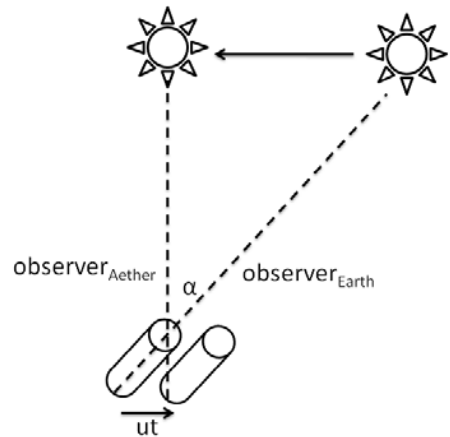

Figure 3. The rays of starlight reckoned by an imaginary observer at rest with respect to a stationary aether and a real observer at rest with respect to the Earth. The imaginary observer sees the "present" position of the star as if the image formed instantaneously and simultaneously with the emission of light while the real observer sees the "past" position of the star that results from the progressive propagation of light. In navigational terms, the "present" position of the star is analogous to the true position reckoned with the aid of calculation and the "past" position of the star is analogous to the apparent position obtained solely with instruments.

In order to visualize the observation of a moving star from the perspective of a stationary observer on Earth, consider a ray of starlight that comes from a star in its "past" position and strikes a glass prism perpendicular to the surface as observed in the inertial frame of the Earth (Figure 4). In this scenario, the image observed in the here and now is *not* formed simultaneously with the image-forming light emitted by the object but only after a period of time necessitated by the progressive motion of light travelling at a finite speed. The ray of starlight emitted by the star is equivalent to the angular wave vector of starlight and the wave fronts that make up the starlight coming from the star in its "past" position are depicted by dotted lines perpendicular to the angular wave vector. An observer at rest with the Earth would point the telescope at the "past" position of the star when it emitted the light seen as an image. The observer could then calculate the "present" position of the star using the angle of aberration. Since the starlight would strike the prism perpendicular to the surface of the prism, the position of the image would be the same with or without the prism, and the angle of aberration ($\alpha$), which is the angle made between the "past" and "present" positions of the star would be the same with or without the prism.

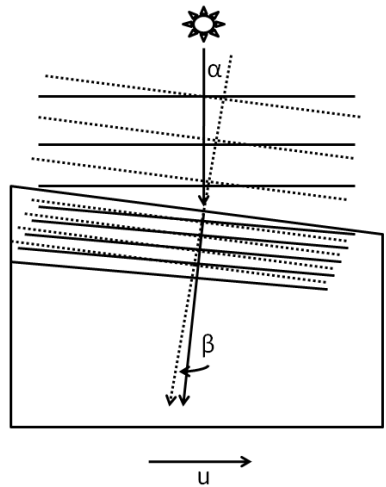

Figure 4. The expected results of introducing a glass prism in front of a telescope on the position of a star observed as a result of the movement of the Earth relative to the fixed stars. A real observer at rest with the Earth who makes the assumption that the image is not observed simultaneously with the emission of the light that forms the image would not point the telescope at the "present" position of the star, but at the position in which the star was in the past when it emitted the light seen as an image of the star. The angle of aberration, which describes the angle made between the "past" and "present" positions of the star would be α. Since the starlight would enter the telescope perpendicular to the surface, this observer would not observe a change in angle of aberration as a result of the introduction of a glass prism. By contrast, an imaginary observer at rest with the stationary aether and who expected the image to be formed instantaneously and simultaneously with the emission of the light that would form the image, would predict that, in the presence of the glass prism, the angle of aberration would be β instead of α, and that the value of β would depend on the velocity of the Earth through the aether. The star is shown in the "present" position.

     Fresnel then considered the observation of starlight from the perspective of an imaginary observer in a reference frame at rest with the stationary aether who is watching the Earth and the glass prism move with velocity *u*. If this observer were to consider the image of the star to be formed instantaneously upon the emission of the light that forms the image, then this observer would see the star in its "present" position. Since the real observer on Earth, who is the only observer with access to the telescope, would have tilted the telescope toward the past position of the star, the observer at rest with respect to the stationary aether, who does not have access to the telescope but has "eyes everywhere" at the present instant, would see the starlight coming from the "present" position of the star strike the surface of the telescope at an angle such that the starlight would subsequently strike the eyepiece of the telescope as it is moving forward through the aether. While the observer on Earth sees the "past" position of the star as being co-parallel with the telescope barrel, the observer at rest with the stationary aether will see the "present" position of the star through the moving telescope (Figure 3).

According to the Snell-Descartes Law, the starlight that strikes the glass prism at an angle relative to the perpendicular bends toward the normal within the prism and creates an angle of refraction (β). The Snell-Descartes Law, which was developed for static or instantaneous situations, which ironically amount to the same thing, describes the bending of light crossing an interface between air and glass with the following equation:

$$n_{air} \sin \alpha = n_{glass} \sin \beta \qquad (2)$$

where $n_{air}$ is the refractive index of air and $n_{glass}$ is the refractive index of glass. Since $n_{air}$ is very close to unity, and since the tangent of an angle is a good approximation of the sine of an angle when the angle is small; and since when an angle is small, the tangent of the angle can be approximated by the angle itself as measured in radians, Equation 2 can be written as:

$$\tan \alpha \approx \alpha \approx n_{glass} \tan \beta \approx n_{glass}\, \beta \qquad (3)$$

Since the refractive index of a transparent medium is the ratio of the velocity of light in the vacuum (*c*) to the velocity of light in the transparent medium (*v*), the reckoning of the refractive index of the glass prism depends on the frame of reference of the observer. After applying the Galilean velocity addition law to the Snell-Descartes Law, an observer at rest with respect to the stationary aether would predict that putting a glass prism in front of the telescope would change the angle of refraction of the starlight, and thus the observed angle of aberration would vary with the velocity (*u*) of the prism. For an observer at rest with respect to the aether frame, the refractive index of the glass would be given *prima facie* by:

$$n_{glass-aether\ frame} = \frac{c+u}{v+u} = \frac{c(1+\frac{u}{c})}{v(1+\frac{u}{v})} = n_{glass} \frac{(1+\frac{u}{c})}{(1+\frac{u}{v})} \qquad (4)$$

which differs from the refractive index of glass ($n_{glass}$) measured in the frame of reference of the Earth and glass prism, from where *u* = 0. Thus an observer at rest with respect to the Earth would predict that the angle of refraction produced by a glass prism would not vary with the motion of the Earth, while an observer at rest with a stationary aether would predict that the angle of refraction produced by a glass prism would vary with the motion of the Earth, and as a result, add to or subtract from the angle of aberration determined without a glass prism.

Since Arago discovered that the calculated angle of aberration was not influenced by the presence of a refractive medium, Fresnel devised a theory that would quantitatively explain Arago's null result for an observer in any frame of reference. Such a theory would also have to allow for stellar aberration. Fresnel formulated a theory by finding a mechanism that would only come into play when the refractive index of the medium was significantly greater than unity and then it would compensate for the bending of light demanded by the Snell-Descartes Law.

From the point of view of the mechanical wave theory of light, the frame-invariant form of the law of refraction must take into consideration the propagation of light with respect to the stationary aether, which according to the theory, determines the speed of light. By using an analogy consistent with the analysis of vibrating elastic strings and the mechanical wave theory of sound, Fresnel postulated that the square of the velocity of light through any medium was proportional to the density of the aether in that medium. However, if all of the aether contained in the glass prism moved at the same velocity as the prism, the refraction of light predicted by the Snell-Descartes Law would be overcompensated. On the other hand, if the aether were perfectly static, the refraction of light predicted by the Snell-Descartes Law would be totally uncompensated. Searching for middle ground, Fresnel postulated that only a portion of the aether in the glass prism was carried along by it when it moved; or equivalently; the aether within the glass prism traveled at velocity $u\phi$, where $\phi$ described the proportion of the aether that would be necessary to be carried along with the glass prism in order to compensate perfectly for the refraction of light predicted by the Snell-Descartes Law.

Again consider a ray of starlight striking the top of a glass prism at an angle ($\alpha$) measured relative to the line perpendicular to the surface as shown in Figures 4 and 5. This ray is seen from the perspective of an observer at rest with respect to a stationary aether and who assumes that the image is formed instantaneously with the emission of light that forms the image. The angle ($\beta$ = BAC) is the angle made by the perpendicular (dotted line) and the angular wave vector (solid line) that describes, from the perspective of an observer at rest with respect to the stationary aether, the instantaneous propagation of light from "present" position of the star. It is clear from Figure 5 that the angle of refraction ($\beta$ = BAC), which is predicted from the Snell-Descartes Law, is smaller than the angle of incidence, which is equal to the aberration angle ($\alpha$ = BAD) reckoned by an observer at rest with respect to the Earth. Angle CAD represents the magnitude that the apparent wave vector must be rotated within the glass prism in the direction the glass prism moves through the aether in order for an observer at rest with respect to the stationary aether to reckon an aberration angle that is independent of the presence of the glass prism. Substituting Equation 1 into Equation 3, we get:

$$\frac{u}{c} \approx n_{glass} \tan \beta \tag{5}$$

If $udt$ represents the distance ($BD$) the glass prism moves through the stationary aether during a given time period ($dt$), $u\phi dt$ represents the distance ($CD$) the aether carried by the glass prism moves during the same time period, and $\frac{c}{n_{glass}} dt$ represents the distance ($AC$) the light propagates through the aether in the glass prism during the same time period. Then, assuming that ABC approximates a right angle, the tangent of the angle of refraction will be given by:

$$\tan \beta \approx \frac{BC}{AC} = \frac{BD-CD}{AC} = \frac{udt - u\phi dt}{\frac{c}{n_{glass}} dt} = \frac{u - u\phi}{\frac{c}{n_{glass}}} = \frac{u n_{glass}}{c}(1 - \phi) \tag{6}$$

Substituting Equation 6 into Equation 5, we get:

$$\frac{u}{c} \approx \frac{u}{c} n_{glass}^2 (1 - \phi) \tag{7}$$

Solving Equation 7 for $\phi$, we find:

$$\phi \approx 1 - \frac{1}{n_{glass}^2} \tag{8}$$

where $\left(1 - \frac{1}{n_{glass}^2}\right)$ is known as Fresnel's dragging coefficient, drag coefficient, partial dragging coefficient, convection coefficient, coefficient of entrainment, or coefficient of entwinement. It represents the portion of aether carried along by the transparent medium or alternatively the portion of the velocity of the transparent medium transmitted to the aether that is necessary to compensate for the refraction of light from the "present" position of the star predicted by the Snell-Descartes Law. Consequently, the Fresnel drag coefficient explains why the angle of aberration is the same, whether a glass prism is placed in front of a telescope or not, to an observer in any frame of reference, including an imaginary observer who is at rest with respect to a stationary aether, who assumes the instantaneous propagation of light and the simultaneity of light emission and image formation, and for whom $u\phi = u\left(1 - \frac{1}{n_{glass}^2}\right)$; and an observer, who is at rest with respect to the glass prism and who does not assume simultaneity and thus points the telescope at the "past" position of the star, and for whom $u\phi = 0\left(1 - \frac{1}{n_{glass}^2}\right) = 0$. The fact that $u\phi$ also vanishes when the refractive index approaches unity allows for the observation of stellar aberration, in terms of the mechanical wave theory, and in practice.

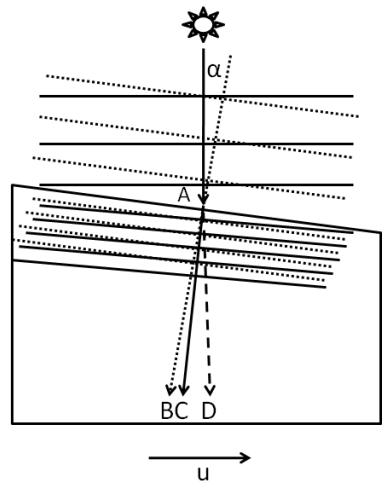

Figure 5. The star is shown in the "present" position. Ray AB describes the ray of starlight predicted by an observer moving with the Earth and at rest with respect to the telescope in the presence or absence of refraction. Ray AC describes the ray of starlight predicted by an imaginary observer at rest with respect to the stationary aether and who assumes that the image is formed instantaneously and simultaneously with the emission of the light that would form the image, in the presence of refraction. This observer would predict that the star would appear to be displaced from its "present" position by refraction. However, observation shows that, from the perspective of an imaginary observer at rest with respect to the stationary aether, the starlight follows ray AD, which is the ray of starlight that would be predicted by an observer at rest with respect to the stationary aether in the absence of refraction. Rays AC and AD are clearly different, yet observation shows that the introduction of a refracting prism has no effect on the angle of aberration. Consequently, Fresnel introduced the dragging coefficient to compensate for the refraction by the glass prism and pull the refracted light that would have followed ray AC so that it would follow ray AD. By introducing the Fresnel drag coefficient, Fresnel was able to reconcile the mutually incompatible requirements of the aether and make the law of stellar aberration and the Snell-Descartes Law laws of physics that were invariant for observers in any frame of reference, including the imaginary observer at rest with respect to the stationary aether who assumes the instantaneous propagation of light and the simultaneity of light emission and image formation.

Above and beyond the assumption of simultaneity held by the imaginary observer at rest with respect to the imaginary aether, is Fresnel's tacit assumption that the glass prism has only a single refractive index that is invariant for all temperatures and for all wavelengths of light, and that only the component of the angular wave vector that is parallel to the velocity of the prism is affected by the motion of the prism. We can explicitly state these tacit assumptions by indicating the wavelength ($\lambda$)- and temperature ($T$)- dependence of the refractive index, and including the cosine of the angle ($\theta$) between the angular wave vector and the velocity vector:

$$\phi \approx \left(1 - \frac{1}{n_{glass(\lambda,T)}^2}\right) \cos \theta \tag{9}$$

The Fresnel drag coefficient is the transformation factor that compensates for the predictions of the Snell-Descartes Law under static conditions so that together they describe the optics of moving transparent media. That is, the Fresnel drag coefficient ($\phi$) is the transformation necessary for the Snell-Descartes Law to be a physical law that is invariant and thus valid in all inertial frames. For example, if the imaginary observer were at rest with respect to the stationary aether, he or she would reckon the speed ($w$) of light propagating through a prism moving through the aether at velocity $u \cos \theta$ to be:

$$w \approx \frac{c}{n_{glass(\lambda,T)}} \pm u\phi \approx \frac{c}{n_{glass(\lambda,T)}} + u \cos \theta \left(1 - \frac{1}{n_{glass(\lambda,T)}^2}\right) \tag{10}$$

Assuming that an observer at rest with the stationary aether would see the "present" position of a star, the starlight would appear to strike the prism at an angle ($\alpha$) relative to a line perpendicular to the surface of the prism. According to Fresnel, the starlight would be subjected to two concurrent effects—it would be partially dragged in the direction of motion of the prism as it was refracted according to the Snell-Descartes Law. As a result, the angle of aberration would be the same with or without the glass prism. By contrast, for an observer at rest with the prism and telescope, $u$ in Equation 10 along with the term representing the Fresnel dragging coefficient would vanish. Such an observer would see the starlight from the "past" position of the same star strike perpendicular to the prism and consequently the angle of aberration would be the same with or without the prism.

Since $\frac{c}{n_{glass(\lambda,T)}} = v_{glass(\lambda,T)}$, the velocity addition law given in Equation 10, which is applicable for any frame of reference, can be written as:

$$w \approx v_{glass(\lambda,T)} \pm u \cos \theta \left(1 - \frac{v_{glass(\lambda,T)}^2}{c^2}\right) \tag{11}$$

Equation 11 explains, from the point of view of the mechanical wave theory, why the composition of velocities, predicted by Arago based on the corpuscular theory of light, did not conform to Galilean relativity where the velocities would be simply added [27]. Specifically, given the newly-developed tenets of the mechanical wave theory of light, and the perspective of an imaginary observer who is at rest with the proper frame of the stationary aether and who assumes the instantaneous propagation of light and simultaneity of light emission and image formation, it appeared that it was the tenacity or viscoelastic properties of the aether that resulted in a nonlinear velocity addition law.

Fresnel concluded his paper by saying that his theory invoking the partial dragging of the aether should be applicable to the experiment previously proposed by Roger Boscovich concerning the observation of stellar aberration through a telescope filled with water or any other fluid more refractive than air and moving relative to the stationary aether at velocity $u$ (Figure 6).

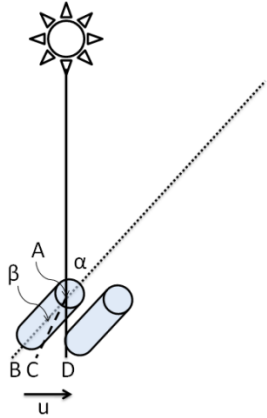

Figure 6. Refraction through a water-filled telescope. The star is shown directly overhead in its "present" position.

Ray AB describes the ray of starlight predicted by an observer moving with the Earth and at rest with respect to the telescope in the presence or absence of water in a telescope, for light that was emitted by a star in the past and took time to propagate through a stationary aether. Since simultaneity is not assumed, the moving star would appear in its "past" position in the telescope. Ray AC describes the ray of starlight predicted by an observer who is at rest with respect to the stationary aether and who assumes simultaneity, in the presence of refraction by water in the telescope. It is predicted that the star would appear to be displaced from its "present" position by refraction. However, observation shows that, from the perspective of an imaginary observer at rest with respect to the stationary aether, the starlight follows ray AD, which is the ray of starlight coming from the star in its "present" position that would be predicted by an observer at rest with respect to the stationary ether, in the absence of refraction. Rays AC and AD are clearly different, yet Fresnel predicted and observation showed that the introduction of water in a telescope had no effect on the angle of aberration. The Fresnel dragging coefficient compensated for the predicted refraction by the water by pulling the refracted light that would have followed ray AC so that it would follow ray AD. By introducing the Fresnel drag coefficient, Fresnel was able to reconcile the mutually incompatible requirements of the aether and make both the law of stellar aberration and the Snell-Descartes Law laws of physics that are invariant for observers in any frame of reference.

The situation shown in Figure 6, like the situation of Arago's prism, can be described by the form of the Snell-Descartes Law used for small angles:

$$\frac{u}{c} \approx n_{water} \tan(BAC) \tag{12}$$

If $udt$ represents the distance ($BD$) the telescope moves through the stationary aether during a given time period ($dt$), $u\phi dt$ represents the distance ($CD$) the aether carried by the water in the telescope moves during the same time period, and $\frac{c}{n_{water}} dt$ represents the distance ($AC$) the light propagates through the water in the telescope during the same time period, then, assuming that ABC approximates a right angle, the tangent of the angle of refraction will be given by:

$$\tan BAC \approx \frac{BC}{AC} = \frac{BD - CD}{AC} = \frac{udt - u\phi dt}{\frac{c}{n_{water}} dt} = \frac{u - u\phi}{\frac{c}{n_{water}}} = \frac{u n_{water}}{c} (1 - \phi) \tag{13}$$

Substituting Equation 13 into Equation 12, we get:

$$\frac{u}{c} \approx \frac{u}{c} n_{water}^2 (1 - \phi) \tag{14}$$

After solving for $\phi$, we find:

$$\phi \approx 1 - \frac{1}{n_{water}^2} \tag{15}$$

Thus Fresnel's drag coefficient again provided the transformation necessary to explain quantitatively why, from any frame of reference, including the frame of reference at rest with respect to a stationary aether in which the instantaneous propagation of light and the simultaneity of light emission and image formation were tacitly assumed, the angle of aberration would be the same in a water-filled telescope as in an air-filled telescope. Fresnel's derivation of the dragging coefficient might not seem all that reliable given that the velocities are referenced to a nonexistent, viscoelastic, stationary, mechanical aether in which an imaginary observer assumes that the image forms instantaneously and simultaneously with the emission of the light that forms the image. Fresnel's derivation might also not be very rigorous [23,24,25,26,30,31,32] given the paucity of equal signs in the derivation; however, this was reasonable and perhaps expected since he was pioneering a new field of wave mechanics. Indeed the descriptive, predictive and explanatory power of Fresnel's wave theory when it came to many optical phenomena, including polarization, interference, diffraction, reflection, refraction as well as stellar aberration led to a near universal acceptance of the mechanical wave theory of light and a reciprocal rejection of Newton's corpuscular theory of light [24,33].

In 1846, George Stokes [34,35] suggested that while Fresnel's complicated solution involving the partial dragging of aether was sufficient to explain stellar aberration, it was not necessary if one took into consideration the friction that would be experienced by the Earth moving through a viscoelastic aether since "*the result would be the same if we supposed the whole of the aether within the earth to move together, the aether entering the earth in front, and being immediately condensed, and issuing from it behind, where it is immediately rarefied, undergoing likewise sudden condensation or rarefaction in passing from one refracting medium to another.*" In 1871, Sir George Airy [36] performed the experiment proposed by Boscovich and showed that the angle of aberration of γ Draconis did not change when the telescope was filled with water instead of air. In his discussion he did not mention whether he thought that the aether was partially dragged by moving bodies as proposed by Fresnel or completely dragged as proposed by Stokes.

Fresnel [29,30] suggested that the aberration of light might be investigated more fruitfully in terrestrial experiments involving a microscope than in astronomical experiments involving a telescope. In order to understand the aberration of light according to the wave theory, Hippolyte Fizeau [37,38,39] designed an interferometer in order to perform a terrestrial experiment that directly tested whether a moving medium did not have any effect on the aether as proposed by Young, completely dragged the aether as proposed by Stokes, or partially dragged the aether as proposed by Fresnel. If the first hypothesis were correct, the velocity of light through a transparent medium would not be affected by the motion of the body at all. If the second hypothesis were correct, the velocity of light through a transparent medium would be augmented by the velocity of the medium, consistent with Galilean relativity. If the third hypothesis were correct, the velocity of light through a transparent medium would be partially augmented by the velocity of the medium, consistent with the Fresnel dragging coefficient and contrary to Galilean relativity.

Fizeau divided a beam of sunlight into two coherent beams with a half-silvered mirror, a converging lens and two slits (Figure 7). One beam propagated through one tube of water and the other beam propagated through a separate and parallel tube. The two beams were then redirected with a converging lens and a mirror so they would enter the tube through which they had not yet propagated. Then the two beams passed through the half-silvered mirror so that their interference pattern could be viewed with a horizontal microscope with an eyepiece micrometer. After observing the position of the interference fringes, Fizeau let the water flow through the tubes such that one beam of light propagated parallel to the movement of the water and the other beam of light propagated antiparallel to the movement of the water. By measuring the shift in the interference fringes, Fizeau could determine whether or not and by how much the aether was dragged with the moving water.

Fizeau's experiment is based on the assumption that that the speed of light ($w$) in a transparent medium moving at velocity ($u$) relative to the laboratory frame is given by the following Equation:

$$w = \frac{c}{n_i} + u\,\phi \tag{16}$$

where $\frac{c}{n_i}$ is the velocity of light propagating through the water when it is at rest relative to the laboratory and $\phi$ is an unknown and dimensionless function to be determined by experiment.

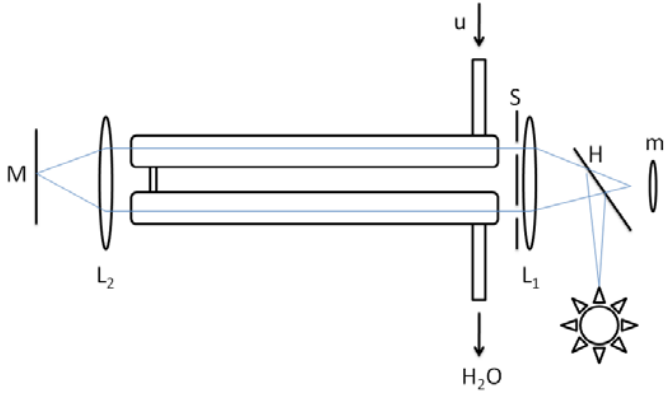

Figure 7. Fizeau's experiment on the propagation of light through moving water. m, microscope with micrometer; H, half-silvered mirror; $L_1$, $L_2$, converging lens; S, slits; M, mirror.

If the aether were stationary, Fizeau would have found that there was no change in the position of the interference pattern and $\phi$ would vanish. If the aether were completely dragged by the moving medium, Fizeau would have found that the interference pattern would have shifted by $\frac{2Lun_i^2}{c\lambda_{source}}$ (see below) and $\phi$ would be unity. Lastly, if the aether were partially dragged by the moving medium, Fizeau would have found an intermediate shift in the interference pattern, and $\phi$ would be between zero and unity, and equal to $(1 - \frac{1}{n_i^2})$.

According to Fizeau, the time ($t_{parallel}$) it would take light to propagate around the interferometer with (parallel to) the motion of the water would be given by:

$$t_{parallel} = \frac{L}{\frac{c}{n_i} + \phi u} \tag{17}$$

and the time ($t_{antiparallel}$) it would take light to propagate around the interferometer against (antiparallel to) the motion of the water would be given by:

$$t_{antiparallel} = \frac{L}{\frac{c}{n_i} - \phi u} \tag{18}$$

where $L$ is the length of the two tubes and $n_i$ is the refractive index of water ($n_i = 1.333$). When the velocity ($u$) of water relative to the laboratory frame equaled zero, then the time difference between the two light beams propagating in the two opposite senses would vanish and

$$t_{parallel} - t_{antiparallel} = 0 \tag{19}$$

But when $u \neq 0$,

$$t_{parallel} - t_{antiparallel} = \frac{L}{\frac{c}{n_i} + \phi u} - \frac{L}{\frac{c}{n_i} - \phi u} \tag{20}$$

and the difference in the optical path length (*OPD*, in m) of light traveling with and against the flow of water would be given by:

$$OPD = \frac{Lc}{\frac{c}{n_i} + \phi u} - \frac{Lc}{\frac{c}{n_i} - \phi u} \tag{21}$$

$$OPD = \frac{Lc(\frac{c}{n_i} - \phi u)}{(\frac{c}{n_i})^2 - (\phi u)^2} - \frac{Lc(\frac{c}{n_i} + \phi u)}{(\frac{c}{n_i})^2 - (\phi u)^2} \tag{22}$$

$$OPD = \frac{-2Lc\phi u}{(\frac{c}{n_i})^2 - (\phi u)^2} \tag{23}$$

Since $\phi u \ll \frac{c}{n_i}$, simplify Equation 23 by neglecting $(\phi u)^2$:

$$OPD = \frac{-2Lc\phi u}{(\frac{c}{n_i})^2} \tag{24}$$

$$OPD = \frac{-2L\phi u n_i^2}{c} \tag{25}$$

If $\phi$ were equal to unity, and the wavelength of the light source in air were $\lambda_{source}$, then the predicted relative fringe shift ($FS = \frac{OPD}{\lambda_{source}}$) would be given by:

$$FS = \frac{OPD}{\lambda_{source}} = \frac{-2Lu n_i^2}{c\lambda_{source}} \tag{26}$$

This result, to first-order accuracy, would be consistent with the formula for the addition of velocities required by Galilean relativity. However, if $\phi$ were equal to $(1 - \frac{1}{n_i^2})$, the predicted relative fringe shift would be given by:

$$FS = \frac{-2Lu(n_i^2 - 1)}{c\lambda_{source}} = \frac{-2Lu n_i^2 (1 - \frac{1}{n_i^2})}{c\lambda_{source}} \tag{27}$$

which was close to Fizeau's experimental results. Consequently, Fizeau concluded that, relative to the laboratory frame, the speed of light propagating through a transparent medium moving at velocity *u* is given by:

$$w = \frac{c}{n_i} + u(1 - \frac{1}{n_i^2}) \tag{28}$$

Fizeau's experiments were repeated by Albert Michelson and Edward Morley [40,41,42] as well as by Pieter Zeeman [43,44,45] with similar results (Table 1) using an optically "brighter" version of the interferometer. The similarity between these experimental results and Fresnel's drag coefficient formula became a watershed event in physics and according to Ludwik Silberstein [46], *"'Agreeing with Fresnel' has become almost a synonym of 'agreeing with experiment.'"* After realizing that the refractive index was a function of wavelength, this meant that the degree that the aether was dragged along with the water would depend on the wavelength. In order to try to understand this complexity, many turned to mathematics to find the exact form of the function that described the wavelength dependence of the predicted fringe shift in Fresnel's drag coefficient [47,48,49,50,51,52,53,54]. Physically, however, a conception of the mechanism of partial aether drag remained obscure.

In order to increase the sensitivity of an experiment designed to measure the speed of light propagating through a moving medium, Martinus Hoek [55] and others [56,57,58,59] redesigned Fizeau's experiment to utilize the speed of the Earth moving around the sun. Hoek designed an interferometer in which light passed through water in one arm and through air in the other (Figure 8). In this way, light traveling in one direction around the interferometer propagated through the water parallel to the motion of the water around the sun and light traveling in the opposite direction propagated antiparallel to the motion of the water around the sun. After finding that the light propagated through the water parallel to the velocity of the Earth at the same speed that it propagated through the water antiparallel to the velocity of the moving Earth, Hoek calculated the function $\phi$ that would compensate for the velocity of the water through the stationary aether and thus explain the vanishing optical path difference between the light propagating in the two directions. Again, the function $\phi$ necessary to give the null result in Hoek's experiment was identical to Fresnel's drag coefficient, further supporting the significance of the Fresnel drag coefficient in understanding the composition of velocities in investigations concerning the optics of moving media.

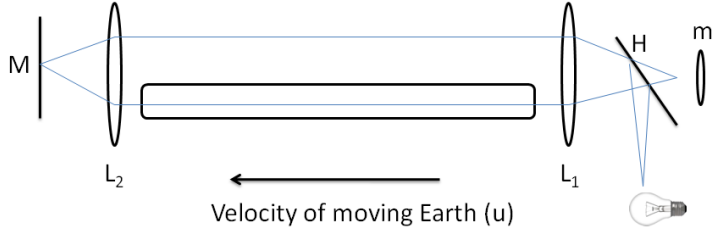

Figure 8. Optical set up of Hoek's experiment on the optics of moving media. $L_1$ and $L_2$ are conversing lenses; H, half-silvered mirror; M, mirror; m, microscope with micrometer.

Hoek calculated the function $\phi$ by assuming that the time required for light to pass through the air in the interferometer arm parallel and antiparallel to the movement of the Earth would be given by $\frac{L}{c+u}$ and $\frac{L}{c-u}$, respectively; and that the time required for light to pass through the water in the interferometer arm parallel and antiparallel to the movement of the Earth would be given by $\frac{L}{\frac{c}{n_i}+u-\phi u}$ and $\frac{L}{\frac{c}{n_i}-u+\phi u}$, respectively. Consequently, the observed null result would be described by the following Equation:

$$\frac{L}{c+u} + \frac{L}{\frac{c}{n_i}-u+\phi u} = \frac{L}{c-u} + \frac{L}{\frac{c}{n_i}+u-\phi u} \tag{29}$$

Putting the denominators in a form ready for simplification using a Taylor expansion we get:

$$\frac{L}{c(1+\frac{v}{c})} + \frac{L}{\frac{c}{n_i}(1-\frac{v}{v'}+\frac{\phi v}{v})} = \frac{L}{c(1-\frac{v}{c})} + \frac{L}{\frac{c}{n_i}(1+\frac{v}{v'}-\frac{\phi v}{v'})} \tag{30}$$

Since $\frac{1}{1+x} = (1-x)$ when $x$ is small and if we neglect terms of second and higher orders from the Taylor expansion, we get:

$$\frac{L(1-\frac{u}{c})}{c} + \frac{L(1+\frac{u}{c}-\frac{\phi u}{c})}{\frac{c}{n_i}} \approx \frac{L(1+\frac{u}{c})}{c} + \frac{L(1-\frac{u}{c}+\frac{\phi u}{c})}{\frac{c}{n_i}} \tag{31}$$

After cancelling like terms and rearranging, we get:

$$\frac{(1-\frac{u}{c})}{c} - \frac{(1+\frac{u}{c})}{c} \approx \frac{(1-\frac{u}{c}+\frac{\phi u}{c})}{\frac{c}{n_i}} - \frac{(1+\frac{u}{c}-\frac{\phi u}{c})}{\frac{c}{n_i}} \tag{32}$$

which can be more simply given as:

$$\frac{1}{c}(1-\frac{u}{c}-1-\frac{u}{c}) \approx \frac{1}{\frac{c}{n_i}}(1-\frac{u}{c}+\frac{\phi u}{c}-1-\frac{u}{c}+\frac{\phi u}{c}) \tag{33}$$

After further simplification we get:

$$\frac{1}{c}(\frac{-2u}{c}) \approx \frac{1}{\frac{c}{n_i}}(\frac{-2u}{\frac{c}{n_i}} + \frac{2\phi u}{\frac{c}{n_i}}) \tag{34}$$

$$\frac{-u}{c^2} \approx \frac{-u}{(\frac{c}{n_i})^2} + \frac{\phi u}{(\frac{c}{n_i})^2} \tag{35}$$

$$\frac{-(\frac{c}{n_i})^2 u}{c^2} \approx -u + \phi u \tag{36}$$

$$\frac{-u}{n_i^2} \approx -u + \phi u \tag{37}$$

which reduced to:

$$\phi \approx (1 - \frac{1}{n_i^2}) \tag{38}$$

which is the Fresnel drag coefficient. Consequently, the formula for the composition of velocities was given by:

$$w = \frac{c}{n_i} + u(1 - \frac{1}{n_i^2}) \tag{39}$$

According to Equation 39, the velocity of light moving through the water and the velocity of the moving water itself are not simply added as would be expected from velocity addition formula according to Galilean relativity [27,60,61,62]. According to Galilean relativity, which was routinely used at a precision limited to the first order with respect to velocity, the results would have been described by the following equation:

$$w = \frac{c}{n_i} + u \tag{40}$$

which would have implied that $\phi$ would have been equal to one, and that the effect of the motion of the water would have been independent of the refractive index of the medium. Clearly Galilean relativity was limited in describing the experimental results obtained from investigating the optics of moving media. Once it was recognized that light waves were electromagnetic and described by Maxwell's wave equation, a need arose to find the correct transformation equations that connected Fresnel's drag coefficient, which was formulated for matter moving through a stationary aether as it pulls its excess aether along with it, with Maxwell's wave equation, which was formulated for matter that was at rest with respect to a stationary aether. Through a series of investigations, Hendrik Lorentz, who greatly admired the work of Fresnel and Maxwell [63,64], set out to find the transformation equations necessary for describing correctly the optics and electrodynamics of moving bodies [65,66,67,68,69,70,71,72,73,74,75]. Lorentz based his work on an assumption of a stationary aether and proposed that Fresnel's drag coefficient could be understood if it were the waves, as opposed to the aether, that were dragged by moving media.

According to the electromagnetic wave theory of light, transparent media were considered to be non-conducting dielectrics and Lorentz assumed that the optical and electrodynamic effects that were observed in moving transparent, dielectric media were mediated by the bound charged particles that composed them. A force exerted on a bound charged particle by the aether would cause the particle to vibrate. Such a vibration would set up a secondary vibration in the aether that would then affect the adjacent charged particles. Since a force transmitted by the aether is not instantaneous and it takes time for a charged particle to accelerate as a result of the force before it re-radiates the force to the aether, various times must be introduced into the equations to determine the value of the force—for example, the time the force is measured and the earlier time the force originated. Lorentz referred to the various times as local times and he considered the local times, not as true times, but only as an aid to the calculation (*mathematische Hilfsmittel*) of events that were not simultaneous. Lorentz initially related the local times with a transformation that was accurate to the first order. Although Lorentz introduced local times in order to describe events that were not simultaneous, he did not relate his local times to the retarded times introduced by Bernhard Riemann, Franz Neumann, Ludvig Lorenz, Alfred Liénard and Emil Wiechert and disregarded by James Clerk Maxwell [23,76,77,78,79,80,81,82,83,84,85,86,87,88,89]. Nevertheless, Lorentz's local times are reminiscent of retarded times and remind us of the two instants in time involved in the emission of light from an object and the production of an image. The formal distinction however between retarded time and local time is that retarded time has a directional component whereas local time does not. The tacit assumption that had been used in characterizing the optics of moving media by Fresnel, who took into consideration reference frames in which the image would be formed either sequentially or simultaneously with the emission of the image-forming light, was made explicit for describing the electrodynamics of moving bodies by formally introducing retarded and local times.

Lorentz's equations were particularly useful for relating the optical and electromagnetic equations applicable to presumed instantaneous and simultaneous events in the stationary aether to the sequential optical and electromagnetic events observed on Earth as it was moving through the stationary aether with a velocity of $30 \frac{km}{s}$. Lorentz's use of local times merely facilitated the physical and mathematical characterization of a real object moving through the aether by introducing a factitious object, stationary with respect to the aether and subject to Maxwell's Equations. The local times were not meant to have any physical significance. Nevertheless, the introduction of local times allowed Lorentz to develop the equations necessary to relate Maxwell's Equations for stationary bodies with the Fresnel drag coefficient. These equations, which are universally known as the Lorentz transformation equations, were able to explain observations and experiments on the optics and electrodynamics of moving bodies, including stellar aberration, Fizeau's experiment, and most notably, the Michelson-Morley experiment. However, mechanistically, according to Lorentz, the Fresnel drag coefficient described the effect of the movement of charged particles being carried by the dielectric on the incoming and outgoing waves in the stationary aether and not the amount of excess aether being dragged by the dielectric moving through a stationary aether.

While Lorentz considered the local times to be nothing but a mathematical tool and distinct from true, general or absolute time, Albert Einstein interpreted the local times as being the true time for each observer traveling at a given velocity relative to the observed system. Consequently, the reckoning of simultaneity became *de facto* a function of velocity and thus relative. *Pari passu*, the proper frame of reference, where events were considered to be simultaneous, switched from the stationary aether, inhabited by an imaginary observer who was all seeing, to the reference frame of a body whose relative velocity ($u$) vanished. Said another way, the moving body in which $u$ was reckoned to be zero, became equivalent to the stationary frame. After Einstein's [90] publication of the Special Theory of Relativity, in which he presented an alternative to Galilean relativity and a new formula for the addition of velocities, Max von Laue [91] reinterpreted the Fresnel drag coefficient in terms of Einstein's formula for the relativistic addition of velocities based on the Special Theory of Relativity and the Lorentz transformation equations. Since the Special Theory of Relativity was based on postulates that did not require an aether, Einstein and von Laue freed scientists to think about the velocity addition formula without the need to consider the aether with its inextricable morass of contradictory requirements. Von Laue's interpretation of the Fresnel drag coefficient became standard physics [92,93,94,95,96,97,98,99,100,101,102,103,104,105,106].

Von Laue [91] derived the Fresnel drag coefficient from the Lorentz transformation equations for space and time. Assume that the light is propagating parallel to the *x*-axis through a transparent medium moving at velocity ($u$) parallel to the *x*-axis. Then the Lorentz transformation equations for comparing space and time in one inertial frame ($x,y,z,t$) compared with another ($x',y',z',t'$) are given by:

$$x = \gamma(x' + ut') \tag{41}$$
$$y = y' \tag{42}$$
$$z = z' \tag{43}$$
$$t = \gamma\left(t' + \frac{ux'}{c^2}\right) \tag{44}$$

The relativistic velocity addition law for an observer at rest with the laboratory frame follows by taking the derivative of Equation 41 with respect to *t*, where the primed inertial frame is the inertial frame of the moving water:

$$w = \frac{dx}{dt} = \gamma\left(\frac{dx'}{dt'}\frac{dt'}{dt} + u\frac{dt'}{dt}\right) \tag{45}$$

Differentiating Equation 44 with respect to $t'$, we get:

$$\frac{dt}{dt'} = \gamma\left(1 + \frac{u}{c^2}\frac{dx'}{dt'}\right) \tag{46}$$

After inverting Equation 46, we get:

$$\frac{dt'}{dt} = \frac{1}{\gamma\left(1 + \frac{u}{c^2}\frac{dx'}{dt'}\right)} \tag{47}$$

Substituting Equation 47 into Equation 45 and writing $\frac{dx'}{dt'}$, the velocity of light propagating through the water as reckoned by an observer in the inertial frame of the water, as $u'_x$, we get

$$w = \frac{u'_x + u}{1 + \frac{uu'_x}{c^2}} \tag{48}$$

Letting $u'_x = \frac{c}{n_i}$, Equation 48 becomes:

$$w = \frac{\frac{c}{n_i} + u}{1 + \frac{u}{n_i c}} \tag{49}$$

After rearranging Equation 49, we get:

$$w = \frac{c}{n_i}\left(1 + \frac{un_i}{c}\right)\left(1 + \frac{u}{n_i c}\right)^{-1} \tag{50}$$

After expanding $\left(1 + \frac{u}{n_i c}\right)^{-1}$ with a Taylor expansion and neglecting terms that are second and higher orders with respect to $\frac{u}{c}$, we get:

$$w \approx \frac{c}{n_i}\left(1 + \frac{un_i}{c}\right)\left(1 - \frac{u}{n_i c}\right) \tag{51}$$

After multiplying the terms in parentheses, we get:

$$w \approx \frac{c}{n_i}\left(1 + \frac{un_i}{c} - \frac{u}{n_i c} - \frac{u^2 n_i}{n_i c^2}\right) \tag{52}$$

After we again neglect any terms that are second order with respect to $\frac{u}{c}$, we get:

$$w \approx \frac{c}{n}\left(1 + \frac{un_i}{c} - \frac{u}{n_i c}\right) \tag{53}$$

Multiply through by $\frac{c}{n_i}$:

$$w \approx \left(\frac{c}{n_i} + u - \frac{u}{n_i^2}\right) \tag{54}$$

Associate the terms that contain $u$:

$$w \approx \frac{c}{n_i} + \left(u - \frac{u}{n_i^2}\right) \tag{55}$$

After factoring out $u$, we get:

$$w \approx \frac{c}{n_i} + u\left(1 - \frac{1}{n_i^2}\right) \tag{56}$$

and we have recovered the Fresnel drag coefficient by assuming that space and time are relative quantities consistent with the Special Theory of Relativity. Von Laue's derivation from the Lorentz transformation equations, as derived by Einstein after taking into consideration the postulates that form the Special Theory of Relativity, indicates that the Fresnel drag coefficient can be divorced from dynamical mechanisms and viewed strictly in terms of relativistic space-time kinematics. According to French [100], *"We have learned also (thanks largely to Einstein) that we should focus on the bare facts of observation, and should not, through our adherence to a particular theory, read more into them than is there."* This is sound scientific advice and consequently, we will not discuss the relativistic phenomena and the velocity addition law in terms of *gedanken* experiments involving space travelers [107,108] and train travelers [109], but only in terms of tested and testable phenomena. Forthwith we refer primarily to the Fizeau experiment and its replicates.

## 2. Results and Discussion

In a previous paper published in this journal [110], one of us developed a new relativistic wave equation, based on the primacy of the Doppler effect that considers the propagation of light between a source and an observer in different inertial frames:

$$\frac{\partial^2 \Psi}{\partial t^2} = c \frac{\omega_{source}}{k_{observer}} \frac{\sqrt{1 + \frac{u \cos\theta}{c}}}{\sqrt{1 - \frac{u \cos\theta}{c}}} \nabla^2 \Psi \tag{57}$$

where $\theta$ is the angle between the velocity vector and the angular wave vector pointing from the source to the observer. When the velocities of the observer and the angular wave vector tend to be parallel, $\cos\theta < 0$ and when the velocities of the observer and the angular wave vector tend to be antiparallel, $\cos\theta > 0$. The above equation only admits the relative velocity ($u$) between the source and the observer and does not admit the introduction of any velocity relative to a nonexistent aether. This relativistic wave equation, which is form-invariant to the second order in all inertial frames, is the equation of motion that describes the properties of light traveling through the vacuum and reckoned by an observer in an inertial frame moving at velocity $u$ relative to the inertial frame of the light source.

This equation is an alternative to Maxwell's wave equation which was developed to describe the propagation of light through a stationary aether in the absence of a source. Equation 57 was also developed independently of the Lorentz transformation equations. Equation 57 can account for the relativity of simultaneity [110] and the observation that the motion of charged particles cannot exceed the speed of light [111,112] without introducing the relativity of space and time. In this paper, we present a generalization of the relativistic Doppler wave equation in order to explain Fizeau's experimental results concerning the propagation of light through moving transparent media.

Equation 57 can be considered as a special case where the refractive index ($n_i$) is unity for light that travels through the vacuum and $n_i = \frac{c}{v_i} = \frac{ck_i}{\omega_i}$. Equation 57 can be generalized for light moving from a source in the vacuum (air) and then through any transparent non-conducting, dielectric medium by explicitly including the refractive index of the dielectric medium through which the light propagates on its way from a source in air to an observer in air[3]. Letting $k_i$ represent the angular wave number of the light in the medium, we get:

$$\frac{\partial^2 \Psi}{\partial t^2} = c \, \frac{n_i \omega_{air-source}}{k_{i-observer}} \frac{\sqrt{1 + \frac{u \cos \theta}{c}}}{\sqrt{1 - \frac{u \cos \theta}{c}}} \nabla^2 \Psi \tag{58}$$

$n_i$ is not included in the Doppler term ($\frac{\sqrt{1 + \frac{u \cos \theta}{c}}}{\sqrt{1 - \frac{u \cos \theta}{c}}}$) since the movement ($u$) of the dielectric medium is limited by the speed of light in the vacuum and not by the speed of light in the transparent medium. A transparent medium moving at a velocity greater than the speed of light in the transparent medium would produce a Mach cone [113] as is seen with Cherenkov radiation [114]. Thus $n_i$ puts a brake on the speed of light in a medium while $c$ puts a break on the speed of the medium.

The following equation is a general plane wave solution to the generalized second order relativistic wave equation given above for the wave in a medium with refractive index $n_i$:

$$\Psi = \Psi_o e^{i(\mathbf{k}_{i-observer} \cdot \mathbf{r} - n_i \omega_{air-source} \frac{\sqrt{1 + \frac{u \cos \theta}{c}}}{\sqrt{1 - \frac{u \cos \theta}{c}}} t)} \tag{59}$$

The general plane wave solution assumes that the direction of **r**, which extends from the source to the observer, is arbitrary but that **k**$_{i\text{-}observer}$ is parallel to **r**. Thus $\theta$ is the angle between the velocity vector and the angular wave vector. We can obtain the form-invariant relativistic dispersion relation by substituting Equation 59 into Equation 58 and taking the spatial and temporal partial derivatives:

$$cn_i \frac{\omega_{source}}{k_{i-observer}} \frac{\sqrt{1 + \frac{u \cos \theta}{c}}}{\sqrt{1 - \frac{u \cos \theta}{c}}} i^2 k_{i-observer}^2 \Psi = i^2 n_i^2 \omega_{air-source}^2 \frac{1 + \frac{u \cos \theta}{c}}{1 - \frac{u \cos \theta}{c}} \Psi \tag{60}$$

---

[3] The appendix describes a wave equation in which the light propagates from the source to the observer entirely through a single, homogenous and isotropic medium with a refractive index of $n_i$.

After canceling like terms, we get:

$$k_{i-observer} = \omega_{air-source} \frac{n_i}{c} \frac{\sqrt{1+\frac{u\cos\theta}{c}}}{\sqrt{1-\frac{u\cos\theta}{c}}} \quad (61)$$

Since $\frac{\omega_{air-source}}{c} = k_{air-source}$, Equation 61 becomes:

$$k_{i-observer} = k_{source}\, n_i \frac{\sqrt{1+\frac{u\cos\theta}{c}}}{\sqrt{1-\frac{u\cos\theta}{c}}} \quad (62)$$

After abbreviating $k_{i-observer}$ by $k_i$, and since $k = \frac{2\pi}{\lambda}$, we can recast Equation 62 in terms of wavelength and we get:

$$\frac{2\pi}{\lambda_i} = \frac{2\pi}{\lambda_{source}}\, n_i \frac{\sqrt{1+\frac{u\cos\theta}{c}}}{\sqrt{1-\frac{u\cos\theta}{c}}} \quad (63)$$

$$\lambda_i = \lambda_{source}\, \frac{1}{n_i} \frac{\sqrt{1-\frac{u\cos\theta}{c}}}{\sqrt{1+\frac{u\cos\theta}{c}}} \quad (64)$$

Equation 64 gives the Doppler-shifted wavelength ($\lambda_i$) of light within a transparent, dielectric medium with refractive index $n_i$ moving at velocity $u$ relative to a source in the vacuum (air) with a vacuum wavelength of $\lambda_{source}$. When the velocities of the observer and the angular wave vector tend to be parallel, $\cos\theta < 0$ and when the velocities of the observer and the angular wave vector tend to be antiparallel, $\cos\theta > 0$.

In Fizeau's experiment, the water and light moved either with (parallel to) or against (antiparallel to) each other making $\cos\theta = \pm 1$. Thus for the two situations, Equation 64 becomes:

$$\lambda_{i-parallel} = \lambda_{source}\, \frac{1}{n_i} \frac{\sqrt{1+\frac{u}{c}}}{\sqrt{1-\frac{u}{c}}} \quad (65)$$

for the parallel case ($u > 0;\ \cos\theta = -1$) and

$$\lambda_{i-antiparallel} = \lambda_{source}\, \frac{1}{n_i} \frac{\sqrt{1-\frac{u}{c}}}{\sqrt{1+\frac{u}{c}}} \quad (66)$$

for the antiparallel case ($u > 0$; $\cos\theta = +1$), where $\lambda_{i-parallel}$ represents the wavelength of light propagating with (parallel to) the flow of water in the inertial frame at rest with respect to the moving water and $\lambda_{i-antiparallel}$ represents the wavelength of light propagating against (antiparallel to) the flow of water in the inertial frame at rest with respect to the moving water. The difference in the wavelengths of light traveling through the medium in the two directions is:

$$\lambda_{i-parallel} - \lambda_{i-antiparallel} = \lambda_{source}\frac{1}{n_i}\frac{\sqrt{1+\frac{u}{c}}}{\sqrt{1-\frac{u}{c}}} - \lambda_{source}\frac{1}{n_i}\frac{\sqrt{1-\frac{u}{c}}}{\sqrt{1+\frac{u}{c}}} \tag{67}$$

The refractive index $n_i$ in the above equations refers to the refractive index of the medium through which the light propagates in between the source and the final observer, which are both in air. The above equation gives the difference in the Doppler shift for a single period of a wave train travelling with and against the flow of water. There are many periods within the tube of flowing medium and in order to calculate the optical path difference between the light waves traveling with (parallel to) and against (antiparallel to) the flow of water, we have to calculate the number of wavelengths ($N$; [115,116,117]) in the medium when $u = 0$. Given that the optical path length (OPL, [118]) in the tubes is $n_i L$, the wavelength of the source light is $\lambda_{source}$, and $\lambda_{source} = n_i \lambda_i$, there are:

$$N = \frac{L}{\lambda_i} = \frac{L}{\frac{\lambda_{source}}{n_i}} = \frac{n_i L}{\lambda_{source}} \tag{68}$$

waves in the tube. Thus the optical path lengths of the light propagating with (parallel to) and against (antiparallel to) the moving water in the tubes are:

$$OPL_{i-parallel} = \frac{n_i L}{\lambda_{source}}\lambda_{source}\frac{1}{n_i}\frac{\sqrt{1+\frac{u}{c}}}{\sqrt{1-\frac{u}{c}}} = L\frac{\sqrt{1+\frac{u}{c}}}{\sqrt{1-\frac{u}{c}}} \tag{69}$$

$$OPL_{i-antiparallel} = \frac{n_i L}{\lambda_{source}}\lambda_{source}\frac{1}{n_i}\frac{\sqrt{1-\frac{u}{c}}}{\sqrt{1+\frac{u}{c}}} = L\frac{\sqrt{1-\frac{u}{c}}}{\sqrt{1+\frac{u}{c}}} \tag{70}$$

And the optical path difference (OPD) between the two propagating beams is:

$$OPD = OPL_{i-parallel} - OPL_{i-antiparallel} = L\frac{\sqrt{1+\frac{u}{c}}}{\sqrt{1-\frac{u}{c}}} - L\frac{\sqrt{1-\frac{u}{c}}}{\sqrt{1+\frac{u}{c}}} \tag{71}$$

Simplify by multiplying each term on the right by 1, where $1 = \frac{\sqrt{1+\frac{u}{c}}}{\sqrt{1+\frac{u}{c}}}$ for the first term on the right and $\frac{\sqrt{1-\frac{u}{c}}}{\sqrt{1-\frac{u}{c}}}$ for the second term on the right:

$$OPD = L \frac{1+\frac{u}{c}}{\sqrt{1-\frac{u^2}{c^2}}} - L \frac{1-\frac{u}{c}}{\sqrt{1-\frac{u^2}{c^2}}} \qquad (72)$$

and simplify so that the equation is accurate to the second order by applying a Taylor expansion:

$$OPD = L \frac{\frac{2u}{c}}{\sqrt{1-\frac{u^2}{c^2}}} = \frac{2Lu}{c} \left(1 + \frac{u^2}{2c^2} + O\left(\frac{u}{c}\right)\right) \qquad (73)$$

Since the speed of the water ($u$) is minuscule compared to the speed of light ($c$), then $\frac{u^2}{c^2}$ and terms of higher orders ($O\left(\frac{u}{c}\right)$) are much less than one. By eliminating the second-order and higher terms, we get:

$$OPD \approx \frac{2Lu}{c} \qquad (74)$$

Equation 74 differs from both Fizeau's equation that utilizes Fresnel's partial drag coefficient equation and the equation of the Special Theory of Relativity based on the relativity of space and time [2]. Moreover, Equation 74 also differs from Equation 26, which is consistent with Galilean relativity. Note that the assumptions used to obtain Equation 74 are not valid at media velocities close to the speed of light, which would be difficult to produce in the laboratory. If such high velocities were attainable, the higher order terms would have to be used. Using the simplified equation that applies when $u \ll c$, the fringe shift ($FS$), which is defined as $\frac{OPD}{\lambda_{source}}$, is given by:

$$FS \approx \frac{2Lu}{c\lambda_{source}} \qquad (75)$$

The fringe shift is proportional to the velocity of the water and the fringe shift vanishes when $u$ vanishes.

In Equation 75, which is based on the primacy of the Doppler effect, the fringe shift is predicted to be independent of the refractive index. This contrasts with predictions made by Fizeau's equation, which directly utilizes the Fresnel drag coefficient, and the Special Theory of Relativity, which states how taking account of the relativity of space and time leads to the Fresnel drag coefficient used in Fizeau's equation [2]. Equation 75 also differs from Equation 26, which was derived using Stokes' assumption of complete aether drag. *Mirabilis dictu*, Table 1 shows that the new relativistic Doppler equation is more accurate than the Fresnel drag coefficient equation and the Special Theory of Relativity in describing the results of experiments performed by Fizeau [39], Michelson and Morley [40], and Zeeman [44].

Table I. A Meta-Analysis of the Experimental and Theoretical Values Obtained for the Effect of a Moving Medium on the Speed of Light Given in Fractions of a Wavelength.

| Length ($L$, in m) | Velocity ($u$, in m/s) | Wavelength ($\lambda_{source}$, in nm) | Experimental Results (Double Displacement, $FS$) | Theoretical (Fresnel Drag Coefficient) | Difference Exp –Theor | Theoretical (Rel Doppler Effect) | Difference Exp –Theor | Reference |
|---|---|---|---|---|---|---|---|---|
| 2.9750 | 7.059 | 526 | 0.4602 | 0.414 | 0.046 | 0.533 | -0.073 | 39 |
| 10 | 1 | 570 | 0.184 | 0.182 | 0.003 | 0.234 | -0.050 | 40 |
| 6.04 | 4.65 | 450 | 0.826 | 0.647 | 0.179 | 0.833 | -0.007 | 44 |
| 6.04 | 4.65 | 458 | 0.808 | 0.636 | 0.172 | 0.818 | -0.010 | 44 |
| 6.04 | 4.65 | 546.1 | 0.656 | 0.533 | 0.123 | 0.686 | -0.030 | 44 |
| 6.04 | 4.65 | 644 | 0.542 | 0.452 | 0.090 | 0.582 | -0.040 | 44 |
| 6.04 | 4.65 | 687 | 0.511 | 0.424 | 0.087 | 0.545 | -0.034 | 44 |
| | | | | | $\bar{x} = +0.100$ | | $\bar{x} = -0.035$ | |
| | | | | | SD = 0.064 | | SD = 0.023 | |

$FS$ = number of fringes in the fringe shift that result from a double displacement (water flowing one way- water flowing the other way). The Special Theory of Relativity and the Fresnel drag coefficient equation for a double displacement is: $FS = \frac{4Lun_i^2}{c\lambda_{source}}(1 - \frac{1}{n_i^2})$, while the relativistic Doppler effect equation for a double displacement is: $FS = \frac{4Lu}{c\lambda_{source}}$. A statistical analysis of the differences between results of experiments and the two theories using a one-tailed t-test for two samples with unequal variances shows that the values given by the new relativistic Doppler effect equation are significantly more accurate than the values given by the Fresnel drag coefficient equation and the Special Theory of Relativity (t = 5.2617, α = 0.0005, n = 7).

In his book entitled, *Relativity. The Special and the General Theory*, Einstein [2] wrote that the Fizeau experiment *"decides in favour of [the velocity addition law] derived from the theory of relativity, and the agreement is, indeed, very exact. According to recent and most excellent measurements by Zeeman, the influence of the velocity of flow v on the propagation of light is represented by [the velocity addition law] to within one per cent."* The fact that the new relativistic Doppler effect describes and explains the results of the Fizeau experiment with more than twice the accuracy of the velocity addition law based on the Special Theory of Relativity is not inconsequential.

In Fizeau's equation, the velocity is relative to the laboratory observer. The fact that any velocity relative to the aether has no place in his equation nor in Equation 75, emphasizes that there should be no need to compensate for the movement of a transparent dielectric medium through the aether with Fresnel's drag coefficient, which requires the refractive index. Using Equation 75 to model Hoek's experiment, there should also be no need to compensate for the movement through the aether since all the components are stationary in the laboratory frame and consequently, $u$ vanishes. Given that $u$ vanishes, there should be no fringe shift and the null result is explained without the need for the Fresnel drag coefficient. Indeed, the aether is superfluous when considering the optics of moving media, and there is no need to consider it as a necessary reference frame for optical experiments.

Given that visible light will only be able to interact with the electrons in the flowing dielectric medium, Equation 75 will only hold when there are sufficient electrons in the dielectric medium to interact with all of the propagating photons in the tube. The number of photons in the tube can be estimated with the following equation that is based on dimensional analysis:

$$\text{number of photons in tube} = \text{PFR}\left(\frac{LA}{c}\right) \tag{76}$$

where PFR is the photon fluence rate (in $\frac{photons}{m^2 s}$), $L$ is the length of the two tubes, $A$ is the cross sectional area of the tube, and $c$ is the speed of light. The number of electrons can be estimated from the following equation, which is based on dimensional analysis:

$$\text{number of electrons in tube} = \rho \frac{LA}{m_b}\left(\frac{e}{b}\right) \tag{77}$$

where $\rho$ is the density of the fluid in the tube, $L$ is the length of the two tubes, $A$ is the cross sectional area of the tube, $m_b$ is the average mass of a baryon, and $\left(\frac{e}{b}\right)$ is the electron to baryon ratio ($\approx$ atomic number to atomic mass ratio) of the fluid in the tube. Consequently, Equation 75 is applicable when

$$\rho \frac{LA}{m_b}\left(\frac{e}{b}\right) > \text{PFR}\left(\frac{LA}{c}\right) \tag{78}$$

Maxwell's relation states that the square of the refractive index is approximately equal to the dielectric constant, which is a measure of the concentration of electrons in a dielectric. This indicates that Equation 75 may not apply to gases with refractive indices close to unity. In Equation 75 as well as all the equations that lead up to Equation 75, we must use the refractive index of the material taking into consideration the temperature in which the experiment is done and the wavelength of the source.

As a result of the Doppler shift, the light that emerges from the water moving in the two directions will have slightly different wavelengths. We can model the interference of these two coherent light waves with slightly different wavelengths from the way they will produce beats [117]. The amplitude ($\Psi$) of the resultant wave will be the sum of the two interfering waves:

$$\Psi = \Psi_o[\cos(\frac{2\pi}{\lambda_{i-parallel}})x] + \Psi_o\cos[(\frac{2\pi}{\lambda_{i-antiparallel}})x] \qquad (79)$$

For convenience, let $\Omega = \frac{1}{2}(\frac{2\pi}{\lambda_{i-parallel}} - \frac{2\pi}{\lambda_{i-antiparallel}})$ and $\Lambda = \frac{1}{2}(\frac{2\pi}{\lambda_{i-parallel}} + \frac{2\pi}{\lambda_{i-antiparallel}})$, then

$$\Psi = 2\Psi_o \cos(\Omega x)\cos(\Lambda x) \qquad (80)$$

The intensity (*I*) of the resultant wave is equal to the square of its amplitude:

$$I = \Psi^2 = 4\Psi_o^2 \cos^2(\Omega x)\cos^2(\Lambda x) \qquad (81)$$

And since $\cos^2(\Omega x)$ is so slowly varying, we can consider it to be a constant, and Equation 81 becomes:

$$I \approx 4\Psi_o^2 \cos^2(\Lambda x) \qquad (82)$$

which for small wavelength shifts will be observed as:

$$I \approx 4\Psi_o^2 \cos^2(\frac{1}{2}(\frac{2\pi}{\lambda_{i-parallel}} + \frac{2\pi}{\lambda_{i-antiparallel}})x) \qquad (83)$$

which can be distinguished from the situation where $u = 0$:

$$I \approx 4\Psi_o^2 \cos^2(\frac{2\pi}{\lambda_{source}}x) \qquad (84)$$

While in the interview with Shankland [1] cited above, Einstein stated that the Michelson-Morley [119] experiment had no influence on his development of the Special Theory of Relativity, pedagogically and historically, the Michelson-Morley experiment has been very important in discussions of the Special Theory of Relativity [120,121,122,123,124,125,126,127,128,129,130,131,132,133,134,135,136,137,138,139, 140,141,142,143]. For this reason, we show that the new relativistic wave equation, based on the primacy of the Doppler effect, also predicts the null effect observed by Michelson and Morley. According to Equation 64, the fringe shift should vanish when there is no aether and the relative velocity between the two light waves propagating in different directions vanishes:

$$OPD = OPL_1 - OPL_2 = L\frac{\sqrt{1+\frac{u\cos\theta}{c}}}{\sqrt{1-\frac{u\cos\theta}{c}}} + L\frac{\sqrt{1+\frac{u\cos\theta}{c}}}{\sqrt{1-\frac{u\cos\theta}{c}}} - L\frac{\sqrt{1-\frac{u\cos\theta}{c}}}{\sqrt{1+\frac{u\cos\theta}{c}}} - L\frac{\sqrt{1+\frac{u\cos\theta}{c}}}{\sqrt{1-\frac{u\cos\theta}{c}}} \qquad (85)$$

In order to describe the geometry of the Michelson-Morley experiment, let $\theta = \pi$ for the first term on the right hand side, let $\theta = 0$ for the second term on the right hand side, let $\theta = \frac{\pi}{2}$ for the third term on the right hand side, and let $\theta = \frac{3\pi}{2}$ for the last term on the right hand side. After calculating the cosines, Equation 85 becomes:

$$OPD = OPL_{parallel} - OPL_{perpendicular} = L\frac{\sqrt{1-\frac{u}{c}}}{\sqrt{1+\frac{u}{c}}} + L\frac{\sqrt{1+\frac{u}{c}}}{\sqrt{1-\frac{u}{c}}} - L\frac{\sqrt{1}}{\sqrt{1}} - L\frac{\sqrt{1}}{\sqrt{1}} \qquad (86)$$

Simplify by multiplying the first two terms on the right hand side by 1, where $1 = \frac{\sqrt{1-\frac{u}{c}}}{\sqrt{1-\frac{u}{c}}}$ for the first term and $\frac{\sqrt{1+\frac{u}{c}}}{\sqrt{1+\frac{u}{c}}}$ for the second term:

$$OPD = OPL_{parallel} - OPL_{perpendicular} = L\frac{1-\frac{u}{c}}{\sqrt{1-\frac{u^2}{c^2}}} + L\frac{1+\frac{u}{c}}{\sqrt{1-\frac{u^2}{c^2}}} - L\frac{\sqrt{1}}{\sqrt{1}} - L\frac{\sqrt{1}}{\sqrt{1}} \qquad (87)$$

Simplify to get:

$$OPD = OPL_{parallel} - OPL_{perpendicular} = \frac{2L}{\sqrt{1-\frac{u^2}{c^2}}} - 2L \qquad (88)$$

Since the velocity of the source relative to the velocity of the interferometer and the observer vanishes, $u = 0$, and Equation 88 becomes:

$$OPD = OPL_{parallel} - OPL_{perpendicular} = 2L - 2L = 0 \qquad (89)$$

And when the OPD vanishes, the fringe shift also vanishes:

$$FS = \frac{OPD}{\lambda_{source}} = 0 \qquad (90)$$

     Thus the new relativistic wave equation based on the primacy of the Doppler effect is consistent with the null result for the Michelson-Morley experiment since Equation 90 holds true independently of the orientation or length of the interferometer arms, and the time during the day or during the year when the measurements are taken.

     Experiments concerning the optics of crystalline and noncrystalline materials show that motion neither induces birefringence ($n_{extraordinary} - n_{ordinary}$) in a material with a single refractive index [145,146,147] nor influences the intrinsic birefringence of a crystal such as calcite or quartz. The lack of effect of motion on birefringence [148] is also consistent with the new relativistic wave equation based on the primacy of the Doppler effect in which the motion-dependent refractive index equals unity because the effect of motion on a single wavelength is inversely proportional to the refractive index (Equation 64) while the number of wavelengths affected by motion is proportional to the refractive index (Equation 68).

     We began this paper discussing the aberration of starlight and will now return to it. The Special Theory of Relativity explains the aberration of starlight in terms of the relativity of space

and time [90,100,149]. If the star and the observer on Earth were stationary, then the components of the velocity of the light in the $x$ and $y$ directions would be given by:

$$c_x = -c \cos \varphi \tag{91}$$

$$c_y = -c \sin \varphi \tag{92}$$

where the light propagates along its wave vector with velocity $c$ (Figure 9).

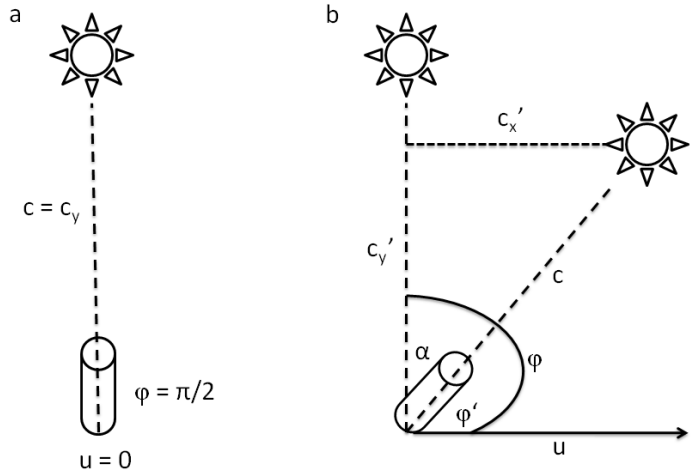

Figure 9. The aberration of starlight. According to the Special Theory of Relativity, the speed of light is invariant. However, as a consequence of the relativity of space and time, the components of the speed of light depend on the relative velocity of the star and the observer. a) $u = 0$; b) $u \neq 0$.

Since the relative velocity of the star and the Earth is $u$, according to the Special Theory of Relativity, the components of the velocity ($c'$) of light in the moving frame of the Earth can be determined by using the Lorentz transformation equations given in Equations 41,42,43, and 44 after transposing the primed and unprimed quantities and after replacing $u$ with $-u$:

$$c_x' = c \cos \varphi' = \frac{c_x - u}{1 - \frac{u c_x}{c^2}} = \frac{-(c \cos \varphi + u)}{1 + \frac{u \cos \varphi}{c}} \tag{93}$$

$$c_y' = c \sin \varphi' = \frac{\frac{c_y}{\gamma}}{1 - \frac{u c_x}{c^2}} = \frac{-c \sin \varphi}{\gamma(1 + \frac{u \cos \varphi}{c})} \tag{94}$$

Again, the observer on Earth reckons that the light propagates along its wave vector with velocity $c$ even though the components of the velocity of light are different for the moving observer and the stationary observer. For the observer on Earth, who according to the Special Theory of Relativity reckons that the vertical and horizontal components of the velocity of light are not equal, the apparent angle ($\varphi'$) of the star relative to the ecliptic will be:

$$\cos \varphi' = -\frac{c_x'}{c} = \frac{\cos \varphi + \frac{u}{c}}{1 + \frac{u \cos \varphi}{c}} \quad (95)$$

Simplify Equation 95 by performing a Taylor expansion and by neglecting terms that are second or higher order with respect to $\frac{u}{c}$:

$$\cos \varphi' \approx (\cos \varphi + \frac{u}{c})(1 - \frac{u \cos \varphi}{c}) = \cos \varphi + \frac{u}{c} - \frac{u \cos^2 \varphi}{c} - \frac{u^2 \cos \varphi}{c^2} = \cos \varphi + \frac{u}{c} - \frac{u \cos^2 \varphi}{c} \quad (96)$$

Simplify Equation 96 using the identity: $sin^2 \varphi = 1 - cos^2 \varphi$

$$\cos \varphi' \approx \cos \varphi + \frac{u}{c} sin^2 \varphi \quad (97)$$

Define the angle of aberration ($\alpha$) as the difference in the angle ($\varphi$), which would be reckoned if the relative tangential velocity between the Earth and the star vanished, and the angle ($\varphi'$), which is reckoned when there is a relative tangential velocity between the star and the Earth moving along the ecliptic:

$$\alpha = \varphi - \varphi' \quad (98)$$

After rearranging Equation 98, and using the trigonometric subtraction formula: $cos (x - y) = cos (x) cos (y) + sin (x) sin (y)$, we get:

$$\cos \varphi' \approx \cos (\varphi - \alpha) = \cos \varphi \cos \alpha + \sin \varphi \sin \alpha \quad (99)$$

Since $\alpha$ is very small, $\cos \alpha \approx 1$ and $\sin \alpha \approx \alpha$, and Equation 99 becomes:

$$\cos \varphi' \approx \cos \varphi + \alpha \sin \varphi \quad (100)$$

Substituting Equation 100 into Equation 97, we get:

$$\cos \varphi' \approx \cos \varphi + \alpha \sin \varphi \approx \cos \varphi + \frac{u}{c} sin^2 \varphi \quad (101)$$

After canceling $\cos \varphi$ and $\sin \varphi$, we get:

$$\alpha \approx \frac{u}{c} (\sin \varphi) \quad (102)$$

After canceling $\sin \varphi$, we get:

$$\alpha \approx \frac{u}{c} \sin \varphi \tag{103}$$

For the case where the position of the star would be over head for a stationary observer ($\varphi = \frac{\pi}{2}$ and $\sin \varphi = 1$), the observed angle of aberration for an observer on Earth moving relative to the star would be given by Equation 104:

$$\alpha \approx \frac{u}{c} \tag{104}$$

which gives the actual angle of aberration observed by Bradley. Thus stellar aberration can be explained by the velocity-dependent differences in the *x* and *y* coordinates of space-time posited by the Special Theory of Relativity.

By contrast with the Special Theory of Relativity, which explains stellar aberration on the basis of the relativity of space and time, the observed angle of aberration can also be explained by the new relativistic wave equation, which is based on the primacy of the Doppler effect [110]. If the new relativistic Doppler effect is the basis of stellar aberration, then we should be able to use the new relativistic Doppler effect to compute the angle of aberration simply and directly, and show its dependence on the relative velocity $u$ and the angle of observation $\varphi'$. Indeed the angle of aberration can be obtained simply by taking the angular derivative of the new relativistic Doppler effect coefficient:

$$\alpha = \frac{d}{d\varphi'} \frac{\sqrt{1 - \frac{u \cos \varphi'}{c}}}{\sqrt{1 + \frac{u \cos \varphi'}{c}}} \tag{105}$$

Simplify:

$$\alpha = \frac{d}{d\varphi'} \frac{1 - \frac{u \cos \varphi'}{c}}{\sqrt{1 - \frac{u^2 \cos^2 \varphi'}{c^2}}} \tag{106}$$

After performing a Taylor expansion and neglecting terms higher than the second order with respect to $\frac{u}{c}$, we get:

$$\alpha \approx \frac{d}{d\varphi'} \left[ \left[ 1 - \frac{u \cos \varphi'}{c} \right] \left[ 1 + \frac{u^2 \cos^2 \varphi'}{2c^2} \right] \right] \tag{107}$$

$$\alpha \approx \frac{d}{d\varphi'} \left[ 1 - \frac{u \cos \varphi'}{c} + \frac{u^2 \cos^2 \varphi'}{2c^2} - \frac{u^3 \cos^3 \varphi'}{2c^3} \right] \tag{108}$$

After taking the derivative, we get:

$$\alpha \approx \frac{u \sin \varphi'}{c} - \frac{u^2 \cos \varphi' \sin \varphi'}{c^2} - \frac{3u^3 \cos^2 \varphi' \sin \varphi'}{2c^3} \tag{109}$$

After neglecting terms that are second order or higher with respect to $\frac{u}{c}$, we get:

$$\alpha \approx \frac{u \sin \varphi'}{c} \tag{110}$$

For the case where the position of the star is nearly overhead, $\sin \varphi \approx 1$, and the observed angle of aberration for an observer on Earth moving relative to the star would be given by:

$$\alpha \approx \frac{u}{c} \tag{111}$$

The relationship between the "past" position of the star and the "present" position of the star can be deduced from the new relativistic Doppler effect by making use of the Principle of Least Time, which was developed by Pierre de Fermat in his quest to understand the refraction of light in transparent media [150,151]. René Descartes, in his *Optics* published in 1637, developed the law of refraction by postulating that light moved from point to point in an instant, no matter what the distance between the points, and that the refraction of light by a transparent medium was a consequence of the relative resistance to light of the incident and transmitting media. Descartes considered the harder transparent medium to exert less resistance to the component of light perpendicular to the interface than the softer air, just as a ball would experience less resistance when rolled across a hard table than it would when rolled with the same force across a soft carpet [152]. In contrast to Descartes' theory of the instantaneous transmission of light, Ole Roemer proposed that the variations in the timing of the eclipses of the moons of Jupiter would be intelligible if light traveled with a finite velocity [153]. The conundrum of the two opposing views of the speed of light is evident in Definition II of Newton's [154] *Opticks*, in which he considered the two mutually-exclusive possibilities that light propagated instantaneously and that light propagated in time. Indeed during the seventeenth century, there were no compelling experimental results that could be used to decide whether the speed of light should be treated as infinitely fast so that an image would be formed instantaneously and simultaneously by a source, or whether the speed of light should be considered to propagate from source to observer in a finite and progressive manner so that an image will be formed after the source emits the light.

Going against Descartes himself, Pierre de Fermat, not only considered the speed of light to be finite, but he used the finite speed of light in a given transparent medium ($v_i$) as the basis of his Principle of Least Time to describe, explain, and predict the processes understood by geometrical optics, including reflection and refraction, the very processes Descartes used to demonstrate the success of his *Method*. In order to describe or predict the position of an image using Fermat's Principle, one must construct an integral for each possible ray that propagates over the distance ($s$) from the source to the observer:

$$t = \int_{source}^{observer} \frac{1}{v_i} ds \tag{112}$$

and then find the ray which takes the least time to propagate from the source to the observer. Fermat interpreted the transit time of light in terms of the index of refraction ($n_i$), which he defined as the ratio of the velocity of light in a vacuum ($c$) to the velocity of light in a transparent material ($v_i$).

$$t = \int_{source}^{observer} \frac{1}{c} n_i ds \tag{113}$$

By eliminating the constant that represents the speed of light in a vacuum, the optical path length (*OPL*) can then be defined as:

$$OPL = \int_{source}^{observer} n_i ds \tag{114}$$

where $n_i$ is the refractive index along an infinitesimal distance $ds$.

Fermat's Principle has been useful for understanding phenomena in geometrical optics [155,156,157,158,159,160] and has served as the basis of the Principle of Least Action in mechanics [161,162,163,164].

The phase of a ray of light is an outsider in geometrical optics, but if one considers the angular wave vector to be equivalent to a light ray, then one can consider the duration of time it takes for light to get from the source to the observer in terms of phase; and anything that affects the phase of the angular wave vector can be incorporated in the integral used to calculate the optical path length or the duration of time it takes for light to get from the source to the observer. The coefficient of the new relativistic Doppler effect describes the velocity ($u$) and angular ($\theta$) dependence of the phase of the angular wave vector pointing from the source:

$$\frac{1 + \frac{u \cos \theta}{c}}{\sqrt{1 - \frac{u^2 \cos^2 \theta}{c^2}}} \tag{115}$$

where $2\pi < \theta < \frac{3\pi}{2}$ when the angular wave vector and the relative velocity vector, both with their origins at the star, point more or less in the same direction, and $2\pi > \theta > \frac{3\pi}{2}$ when the angular wave vector and the relative velocity vector, both with their origins at the star, point more or less in the opposite direction.

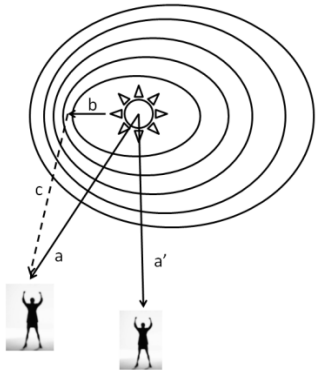

Figure 10. Two observers are equidistant from a star. The total duration of time necessary for light to propagate from a star to the observers who are equidistant from the star and at rest with respect to the star is represented by a and a', where a = a'. When the star is moving relative to the observer, the total duration of time necessary for light to propagate from the star to the observer is given by the difference vector (c) that represents the total duration of time. The magnitude of vector c has physical significance in that it represents the least time to get from the source to the observer, while the direction of vector c points from the observer to the predicted position of the star. If the "past" position of the star is known, the predicted position is the "present" position of the star, and if the "present" position of the star is know, the predicted position is the "past" position.

Both the velocity-independent and the velocity-dependent contribution to the total duration of time it takes for the light to get from the source to the observer that are moving relative to each other at velocity ($u$) can be described and predicted exactly by the following equation (Figure 10):

$$total\ duration = \int_{source}^{observer} \frac{s}{c} \frac{1 + \frac{u\cos\theta}{c}}{\sqrt{1 - \frac{u^2\cos^2\theta}{c^2}}}\ d\theta \tag{116}$$

where *s* is the distance between the source and the observer in the static, velocity-independent case at the initial time. In order to analyze just the velocity-dependent component to the duration, we will subtract the static, velocity-independent component from the total duration:

$$duration = \int_{source}^{observer} \frac{s}{c} \frac{1 + \frac{u\cos\theta}{c}}{\sqrt{1 - \frac{u^2\cos^2\theta}{c^2}}}\ d\theta - \int_{source}^{observer} \frac{s}{c}\ d\theta \tag{117}$$

Since stellar aberration is a first-order phenomenon, we can obtain a first-order equation by reducing the exact solution given above by performing a Taylor expansion and neglecting terms that are higher than second order with respect to $\frac{u}{c}$. After doing so, we get:

$$duration \approx \int_{source}^{observer} \frac{s}{c}\left(1 + \frac{u\cos\theta}{c}\right)\left(1 + \frac{u^2\cos^2\theta}{2c^2}\right)d\theta - \int_{source}^{observer} \frac{s}{c}\ d\theta \tag{118}$$

After multiplying terms, we get:

$$duration \approx \int_{source}^{observer} \frac{s}{c}\left(1 + \frac{u\cos\theta}{c} + \frac{u^2\cos^2\theta}{2c^2} + \frac{u^3\cos^3\theta}{2c^3}\right)d\theta - \int_{source}^{observer} \frac{s}{c}\ d\theta \tag{119}$$

$$duration \approx \int_{source}^{observer} \left(\frac{s}{c} + \frac{su\cos\theta}{c^2} + \frac{su^2\cos^2\theta}{2c^3} + \frac{su^3\cos^3\theta}{2c^4}\right)d\theta - \int_{source}^{observer} \frac{s}{c}\ d\theta \tag{120}$$

After neglecting terms that are second order or higher with respect to $\frac{u}{c}$, we get:

$$duration \approx \int_{source}^{observer} \left(\frac{s}{c} + \frac{su\cos\theta}{c^2}\right)d\theta - \int_{source}^{observer} \frac{s}{c}\ d\theta \tag{121}$$

After removing the terms that are independent of $\theta$ from the inside of the integral, we get:

$$duration \approx \frac{s}{c}\int_{source}^{observer} \left(1 + \frac{u\cos\theta}{c}\right) d\theta \quad - \frac{s}{c}\int_{source}^{observer} d\theta \quad (122)$$

After integrating with respect to $\theta$, we get (within a constant of integration):

$$duration \approx \frac{s}{c}(\theta) + \frac{s}{c}\left(\frac{u\sin\theta}{c}\right) - \frac{s}{c}(\theta) = \frac{su\sin\theta}{c^2} \quad (123)$$

After evaluating the velocity-dependent term for the light propagating in the direction of the observer ($\pi < \theta < \frac{3\pi}{2}$), we see that the duration of time it takes for light to go from a source to an observer decreases by $\frac{su}{c^2}$ compared to when the two are static:

$$duration \approx \left(\frac{su\sin\theta}{c^2}\right) \Big|_{\pi}^{\frac{3\pi}{2}} = -\frac{su}{c^2} \quad (124)$$

By contrast, when evaluating the velocity-dependent term for the light propagating in the direction away from the observer ($2\pi > \theta > \frac{3\pi}{2}$), we see that the duration of time it takes for light to go from a source to an observer increases by $\frac{su}{c^2}$ compared to when the two are static:

$$duration \approx \left(\frac{su\sin\theta}{c^2}\right) \Big|_{\frac{3\pi}{2}}^{2\pi} = \frac{su}{c^2} \quad (125)$$

Equations 124 and 125 give the errors encountered when one assumes that the distance between the source and the observer is minimal and/or the velocity is so small that the moving system can be modeled as a static system. The angles that give the minimal or maximal velocity-dependent change in the duration of time can be conveniently determined by finding the stationary values of the duration obtained from Equation 123:

$$\frac{d\,duration}{d\theta} \approx \frac{d\left(\frac{su\sin\theta}{c^2}\right)}{d\theta} = \frac{su\cos\theta}{c^2} \quad (126)$$

The stationary values of the velocity-dependent change in duration occur when $\theta$ equals either 0 or $\pi$. By taking the derivative of Equation 126, we get:

$$\frac{d\,\frac{su\cos\theta}{c^2}}{d\theta} \approx \frac{-su\sin\theta}{c^2} \quad (127)$$

We will see that, depending on our definition of $u$, the duration of time required for light to get from the source to the observer moving relative to each other is minimized by taking the path where $\theta = \pi$ or $\theta = 0$.

In order to see how Fermat's Principle helps to understand the contribution of the new relativistic Doppler effect to stellar aberration, we will show two ways in which the first-order velocity-dependent contribution to the duration of time it takes light to propagate from the star to the observer can be subtracted from the duration of time calculated under the assumption of stasis. We will do this by positioning the star and the observer two different ways in Cartesian coordinate systems. The first way takes advantage of Richard Feynman's [165] method of reversing the direction of time by starting with the star in the "present" position and then following it as it moves backward in time to its "past" position. The second and more traditional way starts with the star in the "past" position and then follows it as it progresses forward in time to the "present" position. According to Percy Bridgman [166], *"Assuming now that we have our self-contained system of events* [the star emitting light and moving relative to an observer absorbing light]*, we must inquire in detail by what method we assign coordinates to them. This method involves some sort of physical procedure; eventually it must be such that it will give us coordinates in both the stationary and the moving frames of reference. But before we have two coordinate systems we must have one, and issues arise in connection with a single frame of reference which must be solved before we can pass to two."*

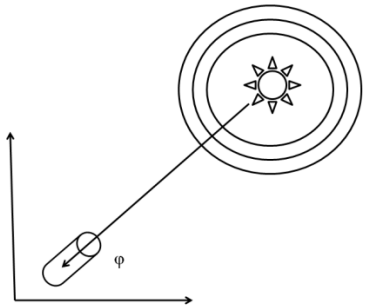

Figure 11. Spherical wave fronts emanating from a star occupying the "present" position when $u = 0$. As Laplace realized centuries ago, the assumption that a moving system can be accurately modeled as a static system is equivalent to assuming that the forces or corpuscles propagate from the source infinitely fast [21]. The instantaneous transmission of force is equivalent to action at a distance.

Consider the static situation where there is no movement ($u = 0$) and where the star occupies a position in the "present" at the instant when the image is seen by an observer. As long as $u = 0$, the light emitted by the star can be represented by spherical and concentric wave fronts (Figure 11). An arc of each wave front is perpendicular to the angular wave vector denoted by the solid line and this wave vector describes the path that takes the least duration of time for light to travel from the source to the observer in a static situation. We can make the model more realistic by taking into consideration both the duration of time predicted for the static situation and the diminution in the duration of time that results from the new relativistic Doppler effect that occurs when there is relative motion between the source and the observer.

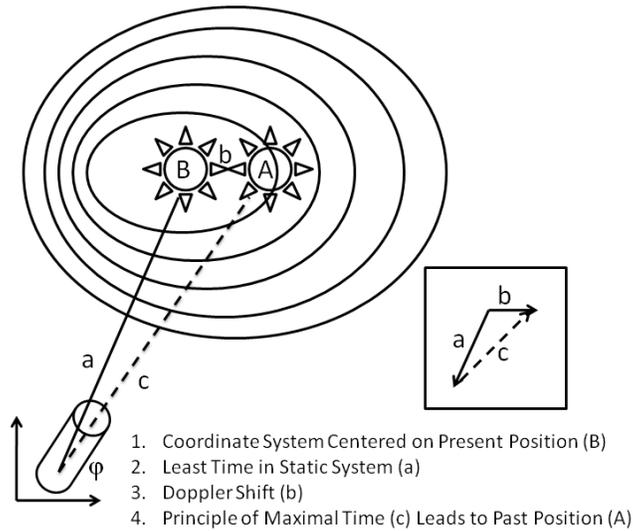

1. Coordinate System Centered on Present Position (B)
2. Least Time in Static System (a)
3. Doppler Shift (b)
4. Principle of Maximal Time (c) Leads to Past Position (A)

Figure 12. $\theta$ is defined for a star moving west, forward in time, or moving east, backwards in time, in a coordinate system centered on B, such that East = 0, North = $\frac{\pi}{2}$, West = $\pi$, and South = $\frac{3\pi}{2}$. When $u = 0$, line a represents the path of least duration of time. When $u \neq 0$, line b represents the first order contribution to decreasing the duration of time it would take light to get from the source to the observer if the system were static. The path that represents the least duration of time is represented to first order by line c. Line c points from the observer in the present to the "past" position of the star. Vectors a, b, and c shown in inset.

      Assuming that the system is static and the star in the "present" position (B) at angle $\varphi$ relative to the horizontal axis the instant the image is formed, we can then retrodict the past by introducing a velocity-dependent term. We do this graphically by drawing the star in the "present" position surrounded by concentric spheroidal waves as described by the new relativistic Doppler effect. The relative velocity of the star has the effect of retarding the phase of the waves between B and A as shown in Figure 12. A star that is moving forward in time with velocity $u$ such that the least time occurs when angle $\theta$ equals $\pi$ can be considered to be a star moving backward in time at velocity $-u$ where the least time occurs at the angle where $\theta$ vanishes. The least time for the velocity-dependent duration is subtracted from the static duration as if they were vectors to get a difference vector. This velocity-dependent time correction, or Dopplerization, which results from the new relativistic Doppler effect, is approximately equivalent to replacing the present time with the retarded time [77]. With an accuracy to the first order, the difference vector points from the observer to the "past" position of the star.

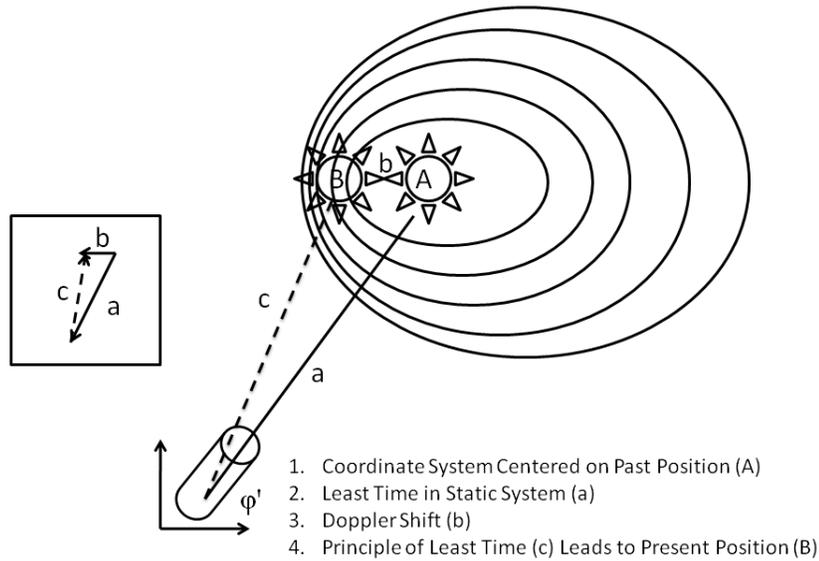

Figure 13. $\theta$ is defined for a star moving west, forward in time, in a coordinate system centered on A, such that East = 0, North = $\frac{\pi}{2}$, West = $\pi$, and South = $\frac{3\pi}{2}$. When $u = 0$, line a represents the path of least duration of time. When $u \neq 0$, line b represents the first order contribution to decreasing the duration of time it would take light to get from the source to the observer if the system were static. The path that represents the least duration of time in a moving system is represented to first order by line c. Line c points from the observer to the "present" position of the star at the instant the image is observed. Vectors a, b and c shown in inset.

We can also model stellar aberration by starting with the assumption that the system is static and the star in the "past" position (A) at angle $\varphi'$ relative to the horizontal axis the instant the image is formed as shown in Figure 13. We can then predict the "present" position of the star by introducing a velocity-dependent term. We do this graphically by drawing the star in the "past" position surrounded by concentric spheroidal waves as described by the new relativistic Doppler effect. The relative velocity of the star has the effect of advancing the phase of the waves between A and B. The minimal stationary value for the velocity-dependent duration is subtracted from the static duration as if they were vectors to get a difference vector. This velocity-dependent time correction, or Dopplerization [167], which results from the Doppler effect, is approximately equivalent to replacing the present time with the advanced time [168]. With an accuracy to the first order, the difference vector points from the observer to the "present" position of the star. The angle between line a and line c is equal to the angle of aberration and can be used to describe the "present" position of the star in the coordinate system of the actual observer, and the "true" position of a star in the standard coordinate system [169,170,171,172].

We have analyzed the dynamic system by first considering the static situations where the position of the star is either in the "present" or the "past" position, and then we added a dynamic term that is first order with respect to $\frac{u}{c}$. When starting from the "present" position, the least total time duration vector points to the "past" position of the star; and when starting from the "past" position, the least total time duration vector points to the "present" position of the star. We have made use of two situations to describe the stationary values of the durations that quantify the "past" position of a star when the "present" position is known and the "present" position of a star when the "past" position is known.

We have provided an account of stellar aberration that incorporates the mathematical world as well as the physical world [173,174]. In doing so we hope that we have provided a mathematically and physically rigorous picture of how stellar aberration can be described and explained by the new relativistic Doppler effect. Paul Dirac [175] wrote that, "*The main object of physical science is not the provision of pictures, but is the formulation of laws governing phenomena and the application of these laws to the discovery of new phenomena. If a picture exists, so much the better….*"

Inspired by the work of Bradley on aberration, Christian Doppler [176,177,178] proposed that, by necessity, relative motion must be taken into consideration in all wave phenomena. Although John Tyndall [179] ended his discussion of the Doppler effect by stating lukewarmly that, "*The ingenuity of the theory is extreme, but its correctness is more than doubtful*." Indeed Hippolyte Fizeau and Ernst Mach [180,181,182] independently predicted that, when one looked at the displacement of spectral lines, the Doppler effect would be useful for determining the radial velocity of stars. Such an astronomical effect was discovered by Sir William Huggins [183,184,185] and later the same effect was discovered independently in terrestrial experiments by Johannes Stark and Antonino Lo Surdo [186,187]. The Doppler effect has proven to be more than fruitful in understanding phenomena ranging from the sound of a moving violin, to the motion of our solar system and galaxy toward the Virgo cluster of galaxies, to the expansion of the universe [188,189,190,191,192,193,194,195,196,197,198,199,200,201,202]. We believe that the Doppler effect will be also useful for understanding stellar aberration and Fizeau's experiment involving the propagation of light through moving water.

Einstein [2,203,204] emphasized the importance of the Fizeau experiment for the development of the Special Theory of Relativity. Realizing the danger of emphasizing formal relationships at the expense of concrete physical reality, we propose that experimentalists could repeat the Fizeau experiment and extend it by using media with different refractive indices. The Special Theory of Relativity, which interprets the quantitative value of the Fresnel drag coefficient in terms of the *"spatio-temporal behavior of systems inhabiting/carrying Minkowski space-time"* [205], predicts that the fringe shift will be best described by the Fresnel coefficient, which is a function of the refractive index. On the other hand, the new relativistic wave equation based on the primacy of the Doppler effect, predicts a different relationship. One could compare the fringe shifts induced by media moving at a given velocity using methanol, which has a refractive index of 1.326 and xylene, which has a refractive index of 1.495 [206,207,208,209]. Since ($n^2_{methanol}$ - 1 = 0.758276) and ($n^2_{xylene}$ - 1 = 1.235025), the Special Theory of Relativity predicts that the fringe shift will be 1.6 times greater with xylene than methanol, while the relativistic Doppler equation predicts that there will be no difference.

We have previously shown that the relativity of simultaneity and the fact that the velocity of charged particles cannot exceed the speed of light do not require the relativity of time posited by the Special Theory of Relativity for their explanation, but can also be explained in terms of the Doppler effect. Here we add the observations on stellar aberration, the optics of moving media exemplified by the Fizeau experiment, and the Michelson-Morley, experiment as additional phenomena that can be explained in terms of the new relativistic wave equation based on the primacy of the Doppler effect, without the need to introduce the velocity-dependent relativity of space and time.

Robert S. Shankland related the following thoughts to Loyd S. Swenson Jr. in an interview in August, 1974 [204]:

*"I think one of the reasons that Einstein was so taken with the Fizeau experiment was that it gave a number. You see, these null experiments, important as they are, are always subject to the question: Well, was there something missing in the experiment that didn't reveal it? Michelson to the end of his days was worried about this point. But when you have a number, and the Fizeau experiment had a number—and another number that Einstein was so interested in was the aberration constant—those not only would be stimuli for a theory, but they would check against a theory in a way that a null experiment could not."*

In light of these words, repeating the Fizeau experiment to test quantitatively the predictions of the Special Theory of Relativity versus those of the new relativistic wave equation based on the primacy of the Doppler effect is crucial. In addition, performing the Fizeau experiment with transparent, non-conducting, dielectric media with differing refractive indices allows for an additional stringent test of the primacy of the relativity of space and time versus the primacy of the new relativistic Doppler effect. Indeed, when discussing the Fizeau experiment, Wallace Kantor [210] wrote, *"It is to be noted as Einstein has suggested that it takes but one experiment in kinematics **on which dynamics is based** to cause a revision of our current understanding and beliefs."*

## 3. Appendix

Equation 58 can be written for the case where the light propagates from the source to the observer entirely through a single medium with a refractive index ($n_i$). In order to transform Equation 58, which models light propagating through air and a transparent medium, we assume that the following conditions represent the properties of light perpendicular to an air-medium interface[4]:

$$\omega_{air-source} = \omega_{i-source} \qquad (A1)$$

$$k_{air-observer} = \frac{k_{i-observer}}{n_i} \qquad (A2)$$

---

[4] By multiplying all terms in Equations A1 and A2 by $\hbar$, we see how energy ($\hbar\omega_{air-source} = \hbar\omega_{i-source}$) and momentum ($\hbar k_{air-observer} = \frac{\hbar k_{i-observer}}{n_i}$) are conserved at an interface.

and replace $c$ and $\frac{n_i \omega_{air-source}}{k_{i-observer}}$ in Equation 58 with $\frac{c}{n_i} = v_i$ and $\frac{\omega_{i-source}}{k_{i-observer}} = v_i$, respectively, to get:

$$\frac{\partial^2 \Psi}{\partial t^2} = \frac{c}{n_i} \frac{\omega_{i-source}}{k_{i-observer}} \frac{\sqrt{1 + \frac{u \cos \theta}{c}}}{\sqrt{1 - \frac{u \cos \theta}{c}}} \nabla^2 \Psi \tag{A3}$$

In a single transparent medium with refractive index $n_i$, when $u = 0$, $\frac{c}{n_i} \frac{\omega_{i-source}}{k_{i-observer}} \frac{\sqrt{1 + \frac{u \cos \theta}{c}}}{\sqrt{1 - \frac{u \cos \theta}{c}}}$ is equal to $\frac{c^2}{n_i^2}$, and Equation A3 becomes:

$$\frac{\partial^2 \Psi}{\partial t^2} = \frac{c^2}{n_i^2} \nabla^2 \Psi = v_i^2 \nabla^2 \Psi \tag{A4}$$

and in a single transparent medium in which $n_i = 1$, when $u = 0$, $\frac{c}{n_i} \frac{\omega_{i-source}}{k_{i-observer}} \frac{\sqrt{1 + \frac{u \cos \theta}{c}}}{\sqrt{1 - \frac{u \cos \theta}{c}}} = c^2$ and Equation A3 becomes:

$$\frac{\partial^2 \Psi}{\partial t^2} = c^2 \nabla^2 \Psi \tag{A5}$$

which is the form of d'Alembert's homogeneous equation obtained by Maxwell for waves propagating through the aether. The general plane wave solution to Equation A3 for the propagation of light from the source to the observer in the same medium is:

$$\Psi = \Psi_o e^{i(k_{i-observer} \cdot \mathbf{r} - \omega_{i-source} \frac{\sqrt{1 + \frac{u \cos \theta}{c}}}{\sqrt{1 - \frac{u \cos \theta}{c}}} t)} \tag{A6}$$

After substituting Equation A6 into Equation A3 and taking the spatial and temporal partial derivatives, we get:

$$\frac{c \omega_{i-source}}{n_i k_{i-observer}} \frac{\sqrt{1 + \frac{u \cos \theta}{c}}}{\sqrt{1 - \frac{u \cos \theta}{c}}} i^2 k_{i-observer}^2 \Psi = i^2 \omega_{i-source}^2 \frac{1 + \frac{u \cos \theta}{c}}{1 - \frac{u \cos \theta}{c}} \Psi \tag{A7}$$

After canceling like terms, we get:

$$\frac{c}{n_i} k_{i-observer} = \omega_{i-source} \frac{\sqrt{1 + \frac{u \cos \theta}{c}}}{\sqrt{1 - \frac{u \cos \theta}{c}}} \tag{A8}$$

Since $\frac{\omega_{i-source}}{c} = \frac{k_{i-source}}{n_i}$, Equation A8 becomes:

$$k_{i-observer} = k_{i-source} \frac{\sqrt{1 + \frac{u \cos \theta}{c}}}{\sqrt{1 - \frac{u \cos \theta}{c}}} \tag{A9}$$

We can recast Equation A9 in terms of wavelength:

$$\lambda_{i-observer} = \lambda_{i-source} \frac{\sqrt{1 - \frac{u \cos \theta}{c}}}{\sqrt{1 + \frac{u \cos \theta}{c}}} \tag{A10}$$

The above equation gives the difference in the Doppler shift for a single period of a wave train travelling with and against the flow of the medium. There are many periods within the two tubes with a total length $L$ containing the flowing medium and in order to calculate the optical path difference between the light waves traveling with (parallel to) and against (antiparallel to) the flow of medium, we have to calculate the number of wavelengths ($N$) in the medium when $u = 0$:

$$N = \frac{L}{\lambda_{i-source}} \tag{A11}$$

Thus the optical path lengths of the light propagating with (parallel to) and against (antiparallel to) the moving medium in the tubes are:

$$OPL_{i-parallel} = \frac{L}{\lambda_{i-source}} \lambda_{i-source} \frac{\sqrt{1 + \frac{u}{c}}}{\sqrt{1 - \frac{u}{c}}} \tag{A12}$$

$$OPL_{i-antiparallel} = \frac{L}{\lambda_{i-source}} \lambda_{i-source} \frac{\sqrt{1 - \frac{u}{c}}}{\sqrt{1 + \frac{u}{c}}} \tag{A13}$$

And the optical path difference (OPD) between the two propagating beams is:

$$OPD = OPL_{i-parallel} - OPL_{i-antiparallel} = L \frac{1 + \frac{u}{c}}{\sqrt{1 - \frac{u^2}{c^2}}} - L \frac{1 - \frac{u}{c}}{\sqrt{1 - \frac{u^2}{c^2}}} \tag{A14}$$

Simplify so that the equation is accurate to the second order by applying a Taylor expansion:

$$OPD = L \frac{\frac{2u}{c}}{\sqrt{1 - \frac{u^2}{c^2}}} = \frac{2Lu}{c} \left(1 + \frac{u^2}{2c^2} + O\left(\frac{u}{c}\right)\right) \tag{A15}$$

Since the speed of the water ($u$) is minuscule compared to the speed of light ($c$), to an accuracy to the first order, Equation A15 becomes:

$$OPD \approx \frac{2Lu}{c} \qquad (A16)$$

The fringe shift (*FS*), which is defined as $\frac{OPD}{\lambda_{i-source}}$, is given by:

$$FS \approx \frac{2Lu}{c\lambda_{i-source}} \qquad (A17)$$

Equation A17, which was derived with the assumption that light propagates from the source to the observer through a single medium, shows that the fringe shift is independent of the refractive index. This is equivalent to the prediction made by Equation 75, which was derived with the assumption that light propagated from the source in air, through moving water, and to an observer in air.